\documentclass[useAMS,usenatbib]{mn2e}

\usepackage[dvipdfmx]{graphicx}
\usepackage{float}

\title[UDF12 Galaxies in Cosmological simulations]
{Physical Properties of UDF12 Galaxies in Cosmological simulations}
\author[Shimizu et al.]
{Ikkoh Shimizu$^{1,2}$\thanks{E-mail: shimizu@astron.s.u-tokyo.ac.jp}, 
Akio K. Inoue$^{1}$, Takashi Okamoto$^{3}$ and Naoki Yoshida$^{4,5}$\\
$^{1}$College of General Education, Osaka Sangyo University, 3-1-1 Nakagaito, Daito, Osaka 574-8530, Japan \\
$^{2}$Department of Astronomy, School of Science, The University of Tokyo, 7-3-1 Hongo, Bunkyo-ku, Tokyo 113-0033 \\
$^{3}$Department of Cosmosciences Graduate School of Science, Hokkaido University, N10 W8, Kitaku, Sapporo, 060-0810, Japan \\
$^{4}$Department of Physics, The University of Tokyo, 7-3-1 Hongo, Bunkyo-ku, Tokyo 113-0033, Japan \\
$^{5}$Kavli Institute for the Physics and Mathematics of the Universe, TODIAS, \\
The University of Tokyo, 5-1-5 Kashiwanoha, Kashiwa, Chiba 277-8583, Japan}
\begin{document}

\date{In original form 2013 May 7}

\pagerange{\pageref{firstpage}--\pageref{lastpage}} \pubyear{2013}

\maketitle

\label{firstpage}

\begin{abstract}
We have performed a large cosmological hydrodynamics simulation tailored
to the deep survey with the Hubble Space Telescope made in 2012, the
so-called UDF12 campaign.
After making a light-cone output, we have applied the same color
selection criteria as the UDF12 campaign to select galaxies from our
simulation, and then, have examined the physical properties of them as a
proxy of the real observed UDF12 galaxies at $z > 7$.
As a result, we find that the halo mass is almost linearly proportional
to the observed ultraviolet (UV) luminosity ($4 \times 10^{11}~{\rm
M_{\odot}}$ at $M_{\rm UV} = -21$).
The dust attenuation and UV slope $\beta$ well correlates with the
observed UV luminosity, which is consistent with observations
quantitatively.
The star formation rate (SFR) is also linearly proportional to the
stellar mass and the specific SFR shows only a weak dependency on the
mass. 
We also find an increasing star formation history with a time-scale of
$\sim100$ Myr in the high-$z$ galaxies.
An average metallicity weighted by the Lyman continuum luminosity
reaches up to $>0.1$ Solar even at $z \sim 10$, suggesting a rapid metal
enrichment.
We also expect $\geq 0.1$ mJy at 350 GHz of the dust thermal emission
from the galaxies with $H_{160} \leq 27$, which can be detectable with
the Atacama Large Milimetre-submilimetre Array.
The galaxies selected by the UDF12 survey contribute to only 52--12\% of
the cosmic SFR density from $z \sim 7$ to $z \sim 10$, respectively.
The James Webb Space Telescope will push the detection fraction up to
77--72\%.
\end{abstract}


\section{Introduction} 
Structure formation in the Universe proceeds from a small scale to 
a large scale through accretion and merging processes of dark matter halos 
in the context of the current concordance $\Lambda$ cold dark matter ($\Lambda$CDM) cosmology. 
Yet baryonic processes making luminous objects such as galaxies are not 
well understood and this is one of the most important questions in the 
modern observational cosmology.
In order to understand the processes of galaxy formation and evolution, 
it is important to look into the site where galaxies are forming in the 
very high-$z$ Universe.
They are intriguing objects as the potential sources causing the cosmic reionization which is also 
one of the hottest topics in the current cosmology research.

High-$z$ galaxy candidates upto $z \sim 7$ have been detected 
by ground-based large telescopes \citep{Ota2008, Ouchi2009, Ouchi2010, Castellano2010, Bowler2012}. 
The redshifts of some of the galaxies are confirmed by spectroscopy 
\citep{Vanzella2011, Ono2012, Shibuya2012, Finkelstein2013}.
However, it is difficult to detect galaxies at $z > 7$ 
using the dropout (Lymam break) method by the ground-based telescopes 
because the Ly$\alpha$ break shifts to the near-infrared (NIR) wavelength 
where the observation is strongly affected by the Earth's atmosphere. 
Therefore, a space telescope which is free from the atmospheric effect 
is needed to detect highest-$z$ galaxies \citep{Yan2004, Bouwens2004}. 
Since the new Wide Field Camera (WFC3) was installed on the Hubble Space 
Telescope (HST) in 2009, its NIR channel has revolutionized 
the search for highest-$z$ galaxies \citep{Oesch2010, McLure2010, Wilkins2010, Labbe2010, Bouwens2010, Bouwens2011}.
In 2012, a new campaign to take the deepest images of the sky ever obtained was conducted with HST/WFC3 
\citep{Ellis2013, McLure2013, Schenker2013, Dunlop2013, Robertson2013, Ono2013, Koekemoer2013, Oesch2013}.
This is called the UDF12 campaign.
A number of candidates of galaxies beyond $z = 7$ have been discovered 
in the UDF12 data \citep{Ellis2013, McLure2013, Schenker2013, Oesch2013}.
The data set is the best one to investigate such highest-$z$ galaxies 
even though their redshifts have not been confirmed by spectroscopy yet 
(see \citet{Inoue2014} about a spectroscopic follow-up strategy with ALMA). 

In order to study the physical properties of high-$z$ galaxies, 
the so-called spectral energy distribution (SED) fitting is often used \citep{Sawicki1998}.
When applying the method to the UDF12 galaxies, however, the narrow coverage of the SEDs causes a problem.
The HST/WFC3 data, in fact, cover only a narrow range of ultraviolet (UV) in the rest-frame of the galaxies.
The galaxies at very high-redshifts are detected only one longest wavelength band 
at about 1.6 $\mu$m in the observer's frame.
Spitzer/IRAC data sometimes help the situation for brightest galaxies, 
but it is limited because of shallowness of the IRAC data.

In the preset paper, we examine the physical properties of 
the UDF12 galaxies by utilizing a cosmological hydrodynamics simulation 
in order to overcome the limitation of the narrow coverage of the SEDs with 
observational data currently available for highest-$z$ galaxies.
Galaxy formation models based on cosmological hydrodynamics simulations are 
now well developed and can reproduce reasonably well statistical quantities 
such as luminosity functions (LFs), stellar mass functions (SMFs), and cosmic star formation history 
\citep[e.g.,][]{Nagamine2010, Salvaterra2011, Salvaterra2013, Shimizu2011, Shimizu2012, Jaacks2012a, Jaacks2012b, Jaacks2013, Biffi2013, Dayal2013}.
In such a simulation, one can select simulated galaxies which would be 
detected by the UDF12 campaign and would satisfy the color selection criteria.
Then, we can investigate the physical properties such as 
halo mass, stellar mass, star formation rate of the selected ``mock-UDF12 galaxies''.
A similar approach has been proposed by \cite{Overzier2013} but their method is 
based on a semi-analytic model of galaxy formation and did not 
discuss the highest-$z$ galaxies sampled by the UDF12 campaign.

In section 2, we describe our numerical simulations and calibration parameters. 
Then, we virtually observe the galaxies produced by our simulation 
through a light-cone output and select simulated galaxies by the same criteria 
as those in the UDF12 survey in section 3.
In section 4, we present physical properties of the selected mock-UDF12 
galaxies in our simulation.
In section 5, we discuss a few implications and predictions.
The final section is devoted for our conclusion.

Throughout this paper, we adopt a $\Lambda$CDM cosmology 
with the matter density $\Omega_{\rm{M}} = 0.27$, 
the cosmological constant $\Omega_{\Lambda} = 0.73$, 
the Hubble constant $h = 0.7$ in the unit of $H_0 = 100 {\rm ~km ~s^{-1} ~Mpc^{-1}}$ and 
the baryon density $\Omega_{\rm B} = 0.046$. 
The matter density fluctuations are normalised by setting
$\sigma_8 = 0.81$ \citep{WMAP}. 
All magnitudes are quoted in the AB system \citep{Oke1990}. 
The assumed initial mass function (IMF) in the observational data and in our simulation 
is always the Salpeter IMF with the mass range of 0.1--100 $M_\odot$.

\section{Cosmological Simulations}
We first describe our cosmological hydrodynamics simulation. 
We calibrate parameters in the code by comparing the SMF observed at $z = 7$. 
Then, we introduce the SED model of the simulated galaxies. 
To do this, we need to adopt a recipe for the dust attenuation which 
has additional parameters.
To calibrate them, we compare the UV LFs in our simulation with the observed
ones at $z = 7, ~8, ~9$ and $z = 10$.
These observational data are compiled from literatures including the UDF12 results.

Our simulation code is based on an updated version of the Tree-PM smoothed 
particle hydrodynamics (SPH) code {\scriptsize GADGET-3} which is a successor 
of Tree-PM SPH code {\scriptsize GADGET-2} \citep{Gadget}. 
We have implemented relevant physical processes such as star formation,
supernova (SN) feedback and chemical enrichment following
\citet{Okamoto2008, Okamoto2009, Okamoto2010}. 
As introduced by \citet{Okamoto2010}, we have assumed that the initial wind 
speed of gas around stellar particles in which SNe occur is proportional
to the local velocity dispersion of dark matter. 
The feedback model is motivated by recent observations of \citet{Martin2005}.  

We employ a total of $2 \times 640^3$ particles for dark matter and gas in a comoving 
volume of $50 h^{-1}{\rm ~Mpc}$ cube.
The mass of a dark matter particle is $3.01 \times 10^7 h^{-1}{\rm M_{\odot}}$ 
and that of a gas particle is initially $6.09 \times 10^6 h^{-1}~{\rm M_{\odot}}$, 
respectively. 
The softening length for the gravitational force is set to be $3~ h^{-1}{\rm kpc}$ in comoving unit. 
Gas particles can spawn star particles when they satisfies a set of standard 
criteria for star formation, and then, their mass is reduced. 
On the other hand, their mass increases when the stellar population evolve and return 
a part of the mass including metal elements into the interstellar medium.
A star particle represents a cluster of stars with a range of stellar masses 
because individual stars can not be resolved in the cosmological hydrodynamic simulation. 
For each snapshot of the simulation, we run a friends-of-friends (FoF) group finder 
with a comoving linking length of 0.2 in units of the mean particle separation
to identify groups of dark matter particles as halos . 
Gas and star particles near dark matter particles which compose a FoF group 
are also regarded as the member of the group. 
Then, we identify gravitationally bound groups of at least 32 total (dark matter+gas+star) particles 
as substructures (subhalos) in each FoF group using SUBFIND algorithm developed by \citet{Springel2001}. 
We regard substructures that contain at least 5 star particles as bona-fide galaxies. 
Above this threshold of the star particle number, the discreteness of the stellar mass begins to disappear. 
We note that the minimum halo mass of our simulated galaxies is around $10^9~{\rm M_{\odot}}$ 
which depends on the relative fraction of dark matter, gas and star particles. 
The minimum stellar mass of our simulated galaxies is around $10^7~{\rm M_{\odot}}$ 
which depends on the age of individual star particle in galaxies. 
Henceforth, we only use galaxies satisfying these conditions for our study. 
This treatment of the simulated galaxies is different from our previous work \citep{Shimizu2011, Shimizu2012} 
and thus affects the shape of the stellar mass function 
at lower-$z$ because the FoF halos may contain many substructures. 
However, the overall effect is not significant at higher-$z$ where we discuss in this paper. 
Some recent theoretical work claims that the effect of the so-called chemical feedback 
(i.e., the transition from PopIII to PopII/PopI stars) that affects the early evolution of very high-$z$ galaxies 
and in particular their metal enrichment history \citep[e.g.,][]{Salvaterra2011, Salvaterra2013, Dayal2013}. 
In this study, however, we do not consider this effect (i.e., we consider only PopII/PopI star population) 
because our simulation resolution is not enough to discuss the effect. 

\begin{figure*}
\includegraphics[width = 160mm]{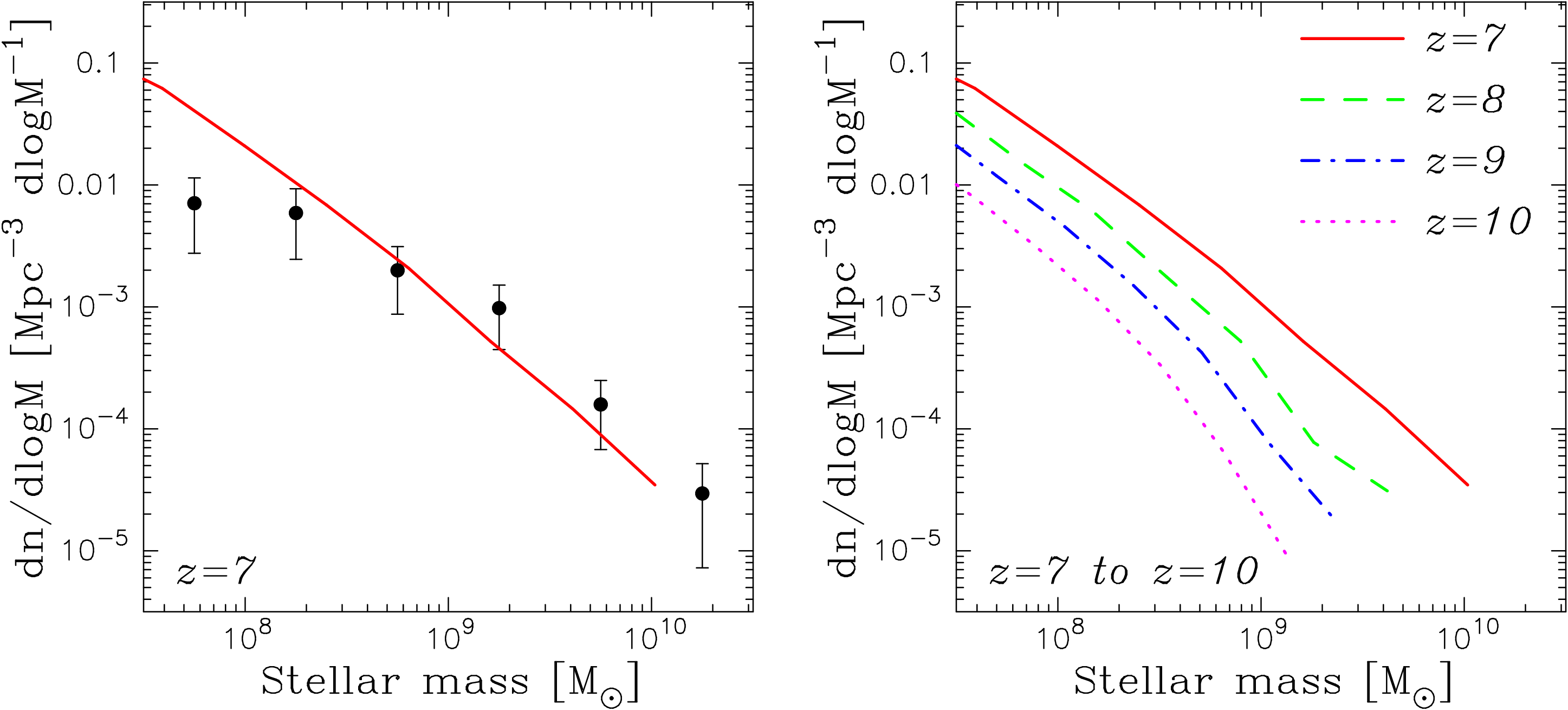}
\caption{The stellar mass functions (SMF) at $z = 7$ (left panel) and higher (right panel). 
The solid, dashed, dash-dotted and dotted lines show the SMFs at $z = 7, ~8, ~9$ and $z = 10$, respectively. 
For $z = 7$, we also plot observational data as the points with error bars 
taken from \citet{Gonzalez2011} in the left panel.} 
\label{FIG1}
\end{figure*}

In order to calibrate parameters in the code such as SN feedback efficiency, 
we compare our model with observations of the stellar mass function (SMF) 
of galaxies, 
assuming the observed SMFs to be complete and correct. 
Fig. \ref{FIG1} represents our SMF at $z = 7$ (left panel) and 
those at $z = 7$ to $z = 10$ (right panel). 
In the left panel, we also plot the observational results 
(the points with error bars) at $z = 7$ taken from \citet{Gonzalez2011}. 
Our model reproduce the observed SMF reasonably well. 
There is an issue of the definition of the stellar mass for low-$z$ galaxies: 
the mass including remnant mass or not \citep{Shimizu2013}. 
However, the effect is negligible at the high-$z$. 

When comparing our simulated galaxies with observed galaxies, 
we calculate the spectral energy distribution (SED) of the simulated galaxies. 
The SED of each star particle which has its own age, metallicity, and mass 
is calculated by using the population synthesis code 
{\scriptsize P\'{E}GASE} \citep{PEGASE}. 
We assume a universal Salpeter initial mass function (IMF) with mass range 
$0.1~{\rm M_{\odot}}$ to $100~{\rm M_{\odot}}$ for all star particles. 
We then sum the SEDs of individual star particles composing a simulated galaxy 
to obtain the total intrinsic SED of the galaxy. 
\citet{Schaerer2008} argued that nebular components (continuum + line) affect 
the rest-frame optical band magnitude for high-$z$ galaxies. 
In this study, however, we consider only the UV range $< 2000~{\rm \AA}$, 
where Ly$\alpha$ emission line is the only important nebular line. 
Some spectroscopic surveys of Ly$\alpha$ emission line at $z > 8$ have been performed, 
but all failed \citep{Brammer2013, Bunker2013, Capak2013, Treu2013}, 
probably because of heavy absorption by intergalactic neutral hydrogen 
before the completion of the cosmic reionization. 
The nebular continuum does not contribute to the SEDs in the UV range 
if the age of galaxies is larger than $50~ {\rm Myr}$ \citep{Inoue2011a}. 
Thus, we do not consider the contribution of nebular components in this study. 

We treat the dust attenuation similarly to \citet{Shimizu2011} and \citet{Shimizu2012}. 
An important difference is that, in this study, we simply adopt the Calzetti law 
\citep{Calzetti2000} rather than using Q values of a specific material and size. 
The optical depth $\tau_{\rm d}$ at $1500~{\rm \AA}$, which is used for the normalization 
of the attenuation law, is calculated by the following equation:
\begin{equation}
\tau_{\rm d} = \frac{3 \Sigma_{\rm d}}{4a_{\rm d} s},
\end{equation}
where $a_{\rm d}$ and $s$ are a typical size and the material density 
of dust grains, respectively. 
We set $a_{\rm d} = 0.1 ~{\rm \mu m}$ and $s = 2.5 ~{\rm g}~{\rm cm}^{-3}$ 
following SNe dust production models \citep{Todini2001, Nozawa2003}. 
Note that any effects of different choices of these values are absorbed in one parameter 
determining the dust surface mass density, $\Sigma_{\rm d}$, as described below.
The surface density is defined by the following equation: 
\begin{equation}
\Sigma_{\rm d} = \frac{M_{\rm d}}{\pi r_{\rm d}^2},
\label{eq:dsmd}
\end{equation}
where $M_{\rm d}$ and $r_{\rm d}$ are the total dust mass and the effective 
radius of the dust distribution in a galaxy, respectively. 
We assume $M_{\rm d}$ and $r_{\rm d}$ to be proportional to the metal mass, $M_{\rm metal}$, 
and the half mass radius, $r_{\rm half}$, of the dark matter distribution in a galaxy, respectively. 
The latter two values are directly calculated in our simulation. 
Therefore, $M_{\rm d}$ and $r_{\rm d}$ are expressed as $e_{\rm Md} M_{\rm metal}$ 
and $e_{\rm rd} r_{\rm half}$, respectively, where $e_{\rm Md}$ and 
$e_{\rm rd}$ are the proportional constants of the dust mass and the effective 
radius, respectively. These two parameters, in fact, can be reduced to one parameter. 
Equation~(\ref{eq:dsmd}) is reduced to
\begin{equation}
\Sigma_{\rm d} = e_{\tau} \frac{M_{\rm metal}}{\pi r_{\rm half}^2}\,,
\end{equation}
where $e_{\tau} = e_{\rm Md} / e_{\rm rd}^2$, which is a global constant 
for all the galaxies in our simulation. 
Then, we calculate the escape probability of UV photons at $1500~{\rm \AA}$ 
($f_{\rm UV}$) using the following equation:
\begin{equation}
f_{\rm UV} = \frac{1 - \delta}{2} (1 + {\rm e}^{- \tau_{\rm d}}) 
               + \frac{\delta}{\tau_{\rm d}} (1 - {\rm e}^{- \tau_{\rm d}}),  
\label{eq:ep}
\end{equation}
where $\delta$ is a parameter whose value is ranging from $0$ to $1$. 
In Equation~(\ref{eq:ep}), a uniform slab where stars and dust are well mixed is sandwiched 
in the middle of two stellar layers, where dust is negligible \citep{Xu1995}. 
The parameter $\delta$ is the fraction of the thickness of the central star+dust 
slab in the total thickness; 
it is the well-mixed dusty slab geometry when $\delta$ is unity, and 
it is the case with a central infinitely thin dusty sheet when $\delta$ is zero.
Such a sandwich geometry is favored to explain the observed relation between 
the UV color and the infrared-to-UV flux ratio of nearby galaxies \citep{Inoue2006}.
In addition, we have found a similar situation in our simulation: 
star particles are sometimes found at the periphery of subhalos and 
the dust (or gas) density at such region is very low. 
The escape probability $f_{\rm UV}$ is saturated at $(1 - \delta) / 2$ 
when $\tau_{\rm d} \to \infty$. 

We calibrate $e_{\tau}$ and $\delta$ so as to reproduce the observed UV luminosity function 
(LF) at $z = 7$, and we keep these values even at higher redshifts. 
For the cosmological simulation in this study, we adopt $e_{\tau} = 3.7$ and $\delta = 0.65$, respectively. 
If we set $e_{\rm Md} = 0.5$, with which we can reproduce the number count of the 
submm galaxies \citep{Shimizu2012}, $e_{\rm rd}$ is uniquely determined as 
$e_{\rm rd} = 0.36$. Interestingly, this value results in the dust distribution radius, 
$r_{\rm d}$, very similar to the half-mass radius of the stellar mass distribution. 
We also apply the IGM absorption for the blue side of $1216 ~{\rm \AA}$ 
following \citet{Madau1995} with an extrapolation for $z > 7$, 
which should not cause a significant uncertainty 
because the spectrum below Ly$\alpha$ is completely damped by the IGM absorption for such high-redshifts. 

\begin{figure*}
\includegraphics[width = 160mm]{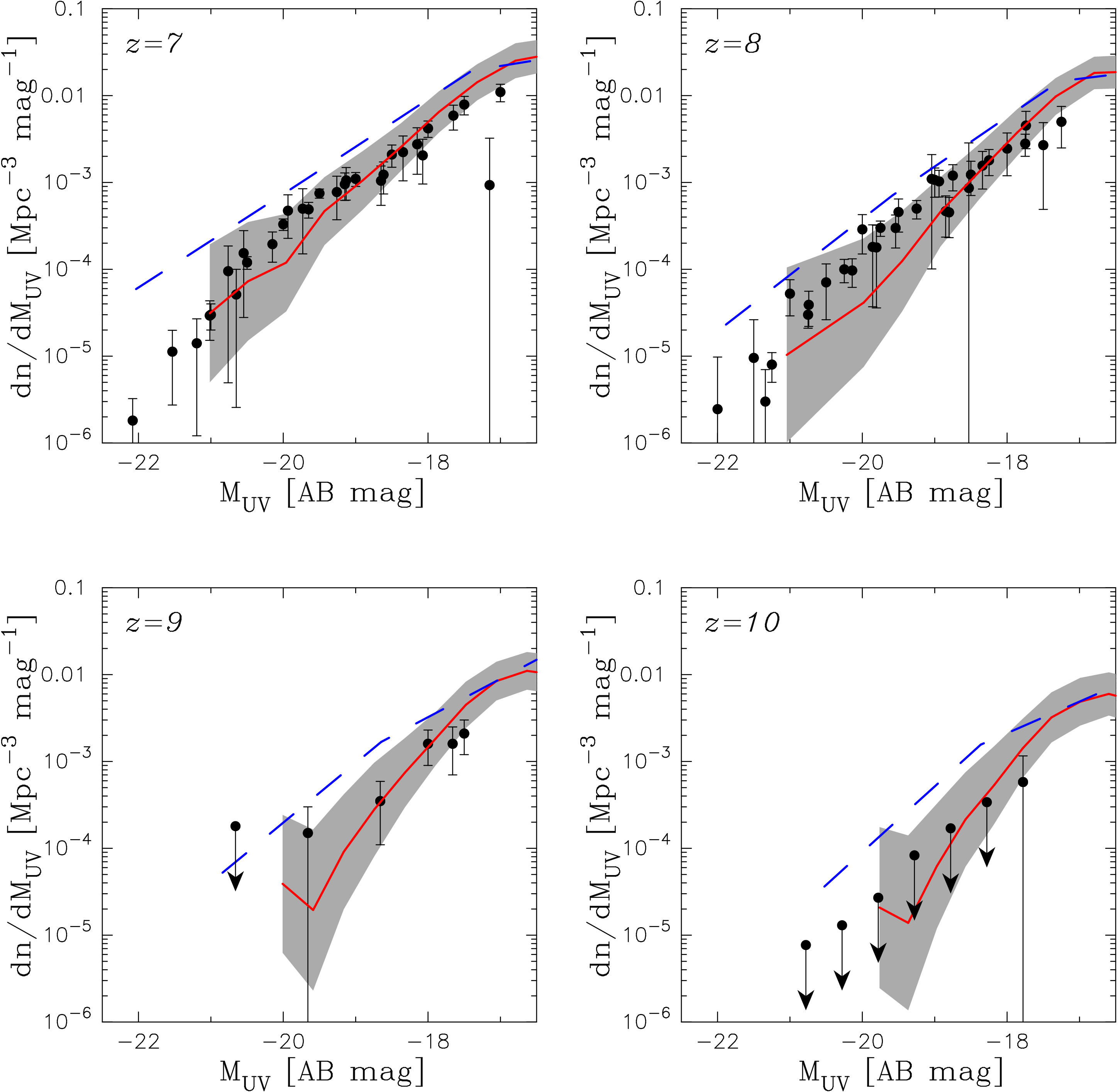}
\caption{The UV luminosity functions (LFs) from $z = 7$ to $z = 10$. 
The solid lines and dashed lines are the UV LFs with and without the dust attenuation, respectively, 
averaged over the entire box of our simulation. 
The shaded regions show the effect of the cosmic variance for a number of the UDF12-sized surveys 
taken from the simulation box in the case with the dust attenuation. 
The points with error-bars or arrows are shown the observational data 
\citep{Robertson2010, Bouwens2011, Bradley2012, Oesch2012, McLure2013, Oesch2013, Schenker2013}. } 
\label{FIG2}
\end{figure*}

Since our simulation box size has the angular size of about $25'$ at $z=7$--10, 
we can have about 130 HST/WFC3 fields-of-view in one side of the box. 
Then, we calculate the UV LFs in the 130 rectangular parallelepiped sub-boxes 
which has the approximately UDF12-sized area in one side. Figure \ref{FIG2} shows 
the average of the UV LFs (solid lines) and their standard deviations (shaded areas) 
of our simulation results at $z = 7, ~8, ~9$ and $z = 10$. 
The points with error-bars represent a compilation of several observational results 
\citep{Robertson2010, Bouwens2011, Bradley2012, Oesch2012, McLure2013, Oesch2013, Schenker2013}. 
Our models reproduce the observed LFs reasonably well within the standard deviations 
which express the cosmic variance of the UDF12-sized pencil beam survey. 
We also show the average UV LFs for the case without dust attenuation as the dashed lines, 
indicating a certain amount of dust attenuation even at $z > 7$. 

\section{Mock Observation}
We need to select galaxies which would be detected and selected in the UDF12 campaign. 
The construction of the mock-UDF12 sample is important for this study.
It should be close similarly to the real observation. 
First, we make a light-cone output from a number of snapshots of the simulation.
Then, we apply the same color selection criteria as for the real UDF12 galaxies. 

In general, the survey depth along the line-of-sight is very wide 
of the Lyman break method using broad-band filters. 
Thus, we can not cover all the survey volume by one snapshot of our simulation. 
In order to directly compare our model with such observations, 
we generate a light-cone output which extends from $z = 6$ to $z = 12$ 
using a number of simulation outputs. 
We coordinate the output redshifts to fill the volume of a light-cone 
from $z = 6$ to $z = 12$ without any gap.
We then randomly shift and rotate each simulation box so that the same objects 
do not appear multiple times on a single line-of-sight at different epochs. 

In order to directly compare our model with observations, 
we calculate broad-band magnitudes of the simulated galaxies on the light-cone 
and choose dropout galaxies with the same manner of the observations.
We use the same set of broad-band filters as the UDF12 which has 
WFC3/IR ($Y_{105}$, $J_{125}$, $JH_{140}$ and $H_{160}$) 
and ACS ($B_{435}$, $V_{606}$, $i_{775}$, $I_{814}$ and $z_{850}$) filters. 
We first apply the detection limit of the UDF12 survey. 
The conditions are $Y_{105} \leq 30$ and $J_{125} \leq 30$ for the $z \sim 7$ selection, 
$J_{125} \leq 30$ and $H_{160} \leq 30$ for the $z \sim 8$ selection and 
$JH_{140} \leq 30$ and $H_{160} \leq 30$ for the $z \sim 9$ and the $z \sim 10$ selections, respectively. 
Then, we apply the color selection criteria for $z \sim 7$ as 
\begin{eqnarray}
z_{850} - Y_{105} &>& 0.7 \nonumber \\
Y_{105} - J_{125} &<& 0.4, 
\label{eq:z7selc}
\end{eqnarray}
and for $z \sim 8$ as 
\begin{eqnarray}
Y_{105} - J_{125} &>& 0.5 \nonumber \\
J_{125} - H_{160} &<& 0.4 ,
\label{eq:z8selc}
\end{eqnarray}
as in \citet{Schenker2013}. 
The color selection criteria for $z \sim9$ are 
\begin{eqnarray}
(Y_{105} + J_{125}) / 2 - JH_{140} &>& 0.75 \nonumber \\
(Y_{105} + J_{125}) / 2 - JH_{140} &>& 0.75 + 1.3 \times (JH_{140} -H_{160}) \nonumber \\
\label{eq:z9selc}
\end{eqnarray}
and for $z \sim10$ are
\begin{eqnarray}
J_{125} - H_{160} &>& 1.2 \nonumber \\
JH_{140} - H_{160} &<& 1.0,  
\label{eq:z10selc}
\end{eqnarray}
as in \citet{Oesch2013}. 
Note that we consider no effect occurring in real observations such as 
observational noise, surface brightness dimming, object detection incompleteness, 
etc., in the selection procedure. Therefore, we have assumed these effects to be 
small. We reserve to examine these effects in future work.

\begin{figure*}
\includegraphics[width = 180mm]{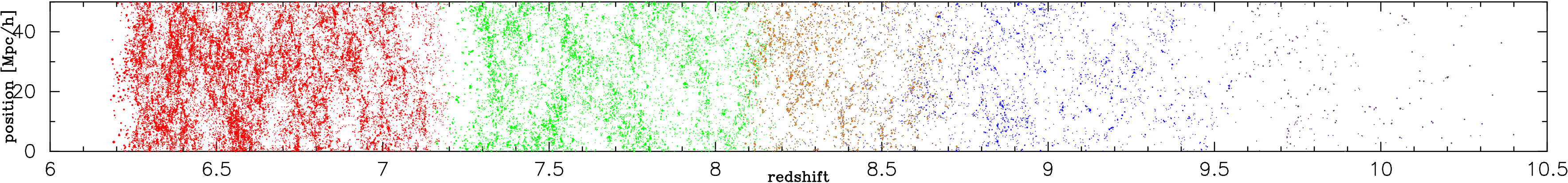}
\caption{The spatial distribution of the simulated galaxies satisfying 
at least one color selection criterion. 
The red, green, blue and magenta points represent the galaxies selected by 
the criteria for $z \sim 7$, 8, 9 and 10, respectively. 
The brown points are the galaxies selected by both criteria for $z \sim 8$ and 9. 
The point size represents $H_{160}$ band magnitude from $H_{160} = 25$ to 30 mag
which is the detection limit of the UDF12 survey. } 
\label{GAL_DIST}
\end{figure*}

Fig. \ref{GAL_DIST} shows the spacial distribution of the color selected simulated 
galaxies. 
The red, green, blue and magenta points represent the simulated galaxies 
selected by the criteria for $z \sim 7, ~8, ~9$ and $z \sim 10$, respectively. 
The brown points are the galaxies doubly selected by the criteria for 
$z \sim 8$ and $z \sim 9$. 
The point size represents $H_{160}$ band magnitude and the largest (smallest) one 
corresponds to $H_{160} = 25~(H_{160} = 30)$. 
We note that almost all simulated galaxies selected by the criteria of $z \sim 10$ 
are also selected by the criteria of $z \sim 9$. 
We find that only a small number of galaxies are selected at $z > 10$ in our simulation. 

\begin{figure}
\includegraphics[width = 90mm]{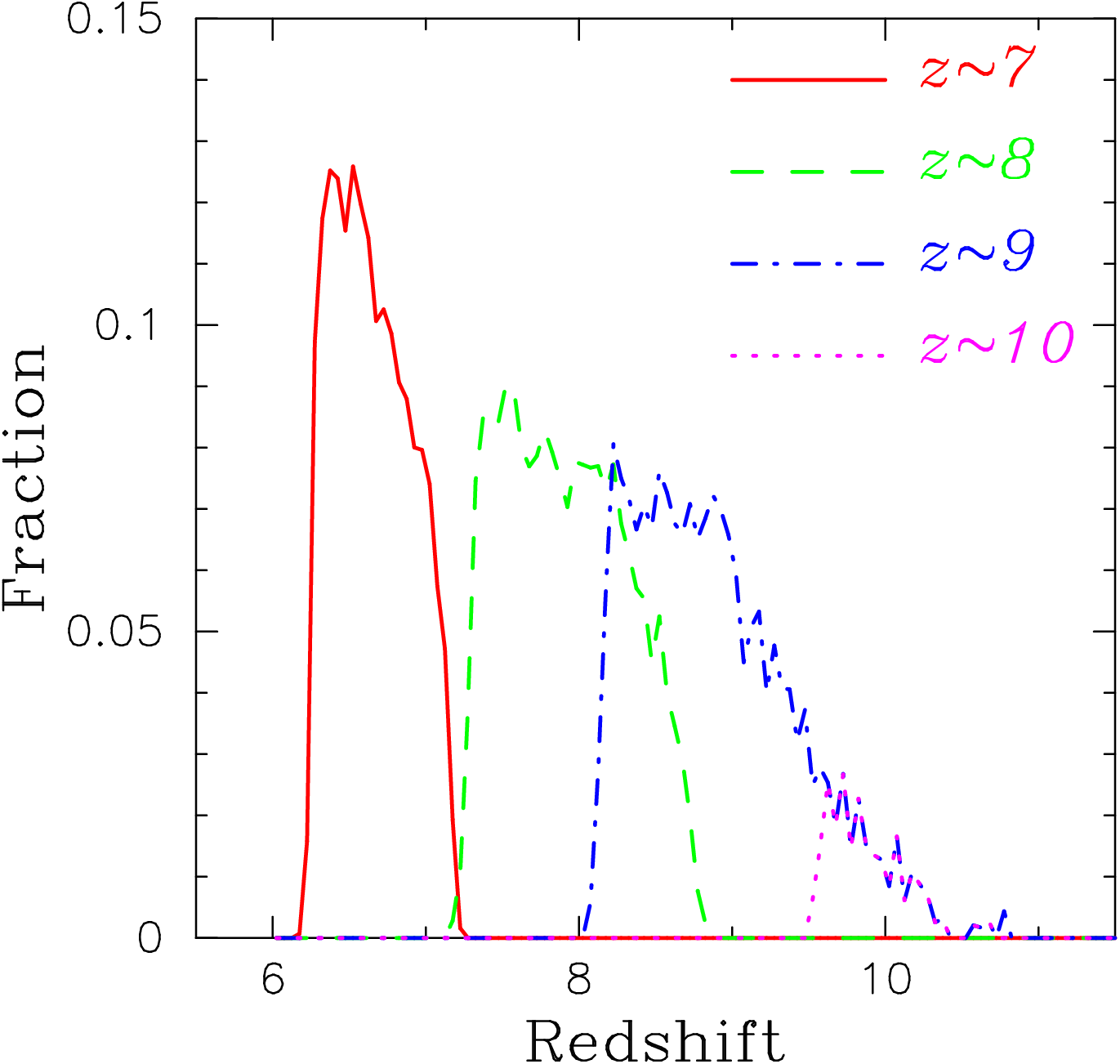}
\caption{The number fraction of 
galaxies brighter than the observational limiting magnitude of 30 mag 
and satisfying the color criteria among all the galaxies resolved in our simulation 
as a function of redshift. The solid, dashed, dash-dotted and dotted lines are the fraction 
for $z \sim 7, ~8, ~9$ and $z \sim 10$ selections, respectively. 
The redshift ranges of galaxies selected are $6.2 \leq z \leq 7.2$ ($z \sim 7$ color selection), 
$7.1 \leq z \leq 8.8$ ($z \sim 8$ color selection), $8.0 \leq z \leq 10.8$ ($z \sim 9$ color selection) 
and $9.5 \leq z \leq 10.6$ ($z \sim 10$ color selection), respectively. } 
\label{SEL_EFFICIENCY}
\end{figure}

We also check the selection efficiency of the color selections. 
Fig. \ref{SEL_EFFICIENCY} presents the number fraction of galaxies 
brighter than the observational limiting magnitude of 30 mag
and satisfying the color criteria among all the galaxies in our simulation 
(see sec.2 for the definition of the galaxies) as a function of redshift. 
The solid, dashed, dash-dotted and dotted lines are the fractions for $z \sim 7, ~8, ~9$ 
and $z \sim 10$ selections, respectively. 
The redshift ranges selected are $6.2 \leq z \leq 7.2$ ($z \sim 7$ color selection), 
$7.1 \leq z \leq 8.8$ ($z \sim 8$ color selection), 
$8.0 \leq z \leq 10.8$ ($z \sim 9$ color selection) 
and $9.5 \leq z \leq 10.6$ ($z \sim 10$ color selection). 
We find that the color selection works reasonably well. 
Next, we note a remarkably small number fraction of the galaxies selected 
as mock-UDF12 galaxies from all galaxies in our simulation. 
Interestingly, the number fraction of the color selected galaxies is less than 
$13 \%$ even at $z \sim 7$ and less than $3 \%$ at $z \sim 10$. 
This means that the galaxies detected in the current deepest survey are still 
just the tip of iceberg and we probably miss almost all star forming galaxies at $z > 7$. 
We revisit this issue later in section 5.3. 

\section{Results}
We have constructed a mock-UDF12 galaxy sample from our simulation 
in the previous section and we are now ready to examine their physical properties.
First, we examine the halo and stellar masses of the mock-UDF12 galaxies.
Second, the dust attenuation and UV slope of the mock-UDF12 galaxies are discussed. 
Third, we look into the correlation between the stellar mass and star formation 
rate (SFR) and specific SFR. 
We also check the cosmic SFR density at $z > 7$. 
Then, the ages and the star formation history of the galaxies are examined.
Finally, the metallicity, morphology and size are investigated. 

\subsection{Halo Mass and Stellar Mass}
\begin{figure*}
\includegraphics[width = 160mm]{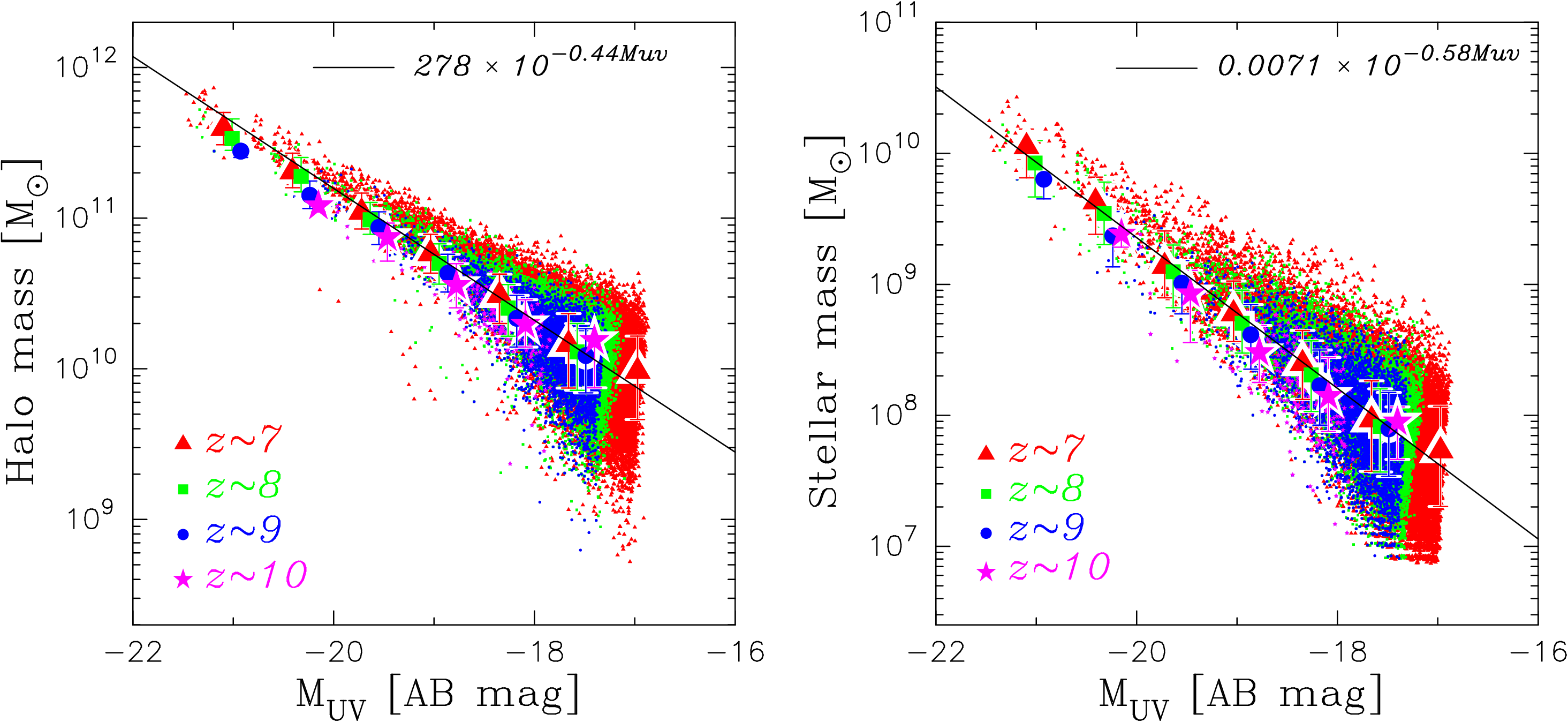}
\caption{The halo mass (left panel) and the stellar mass (right panel) 
as a function of the `observed' absolute UV magnitude calculated directly from 
the observable magnitude without any correction for dust attenuation.
The small triangle, square, circle and star points represent the mock-UDF12 galaxies 
at $z \sim 7, ~8, ~9$ and $z \sim 10$, respectively. 
The large points with error-bars are the median values and the central $68\%$ ranges in halo mass bins. 
The solid lines show the fitting functions and the formulae are noted in the panels. } 
\label{UV_HM_SM}
\end{figure*}

In this subsection, we study the relation between the UV magnitude and 
the halo or stellar mass. 
The left and right panels in Fig. \ref{UV_HM_SM} represent the halo masses and 
the stellar masses of the mock-UDF12 galaxies, respectively, as a function of 
the absolute UV magnitude calculated with the dust attenuation. 
Hereafter, we denote the absolute UV magnitude with the dust attenuation 
as just the absolute UV magnitude or the UV magnitude. 
The small triangle, square, circle and star points represent the each mock-UDF12 galaxy 
at $z \sim 7, ~8, ~9$ and $z \sim 10$, respectively. 
The large points with error-bars are the median values and the central $68\%$ ranges in halo mass bins. 
Interestingly, the relation between the halo or stellar mass 
and the UV magnitude is not significantly different for different redshifts. 
This indicates that the mass--luminosity relation for the mock-UDF12 galaxies 
is independent of the redshift and the mass-luminosity ratio in a certain halo 
mass at different epoch is almost constant.
We here present simple formulae to estimate the halo or stellar mass from the UV magnitude.
Let us assume a functional form as $a \times 10^{b \times M_{\rm UV}}$ where 
$a$ and $b$ are the fitting parameters. 
If $b$ is equal to $-0.4$, then the UV luminosity is linearly proportional 
to the host halo (or stellar) mass. 
For the mock-UDF12 galaxies, we have obtained 
\begin{equation}
\frac{M_{\rm halo}}{M_\odot} = 278 \times 10^{-0.44 M_{\rm UV}}\,,
\label{eq:Mhalo-MUV}
\end{equation} 
for the halo mass relation and 
\begin{equation}
\frac{M_*}{M_\odot} = 0.0071 \times 10^{-0.58 M_{\rm UV}}\,,
\label{eq:Mstellar-MUV}
\end{equation}
for the stellar mass relation. 
Therefore, the UV luminosity is almost linearly proportional to the halo mass for the mock-UDF12 galaxies, 
but the stellar mass scales more strongly with the magnitude. 
Using Equation~(\ref{eq:Mhalo-MUV}) and (\ref{eq:Mstellar-MUV}), we have obtained
\begin{equation}
\frac{M_*}{M_{\rm halo}} = 0.012 \left( \frac{M_{\rm halo}}{10^{11}~{\rm M_{\odot}}} \right)^{0.31}, 
\end{equation}
for the relation between the halo mass and the stellar mass 
when the halo mass is less than $10^{12}$ $M_\odot$. 
Interestingly, this is similar to the results at $z = 7$ and 8 of \citet{Behroozi2013} 
in which they have performed a multi-epoch abundance matching method. 

\subsection{Dust Attenuation}
According to \cite{Dunlop2013}, the UDF12 galaxies at $z \geq  7$ are little  
affected by dust. 
However, as already shown in Fig. \ref{FIG2}, we need a certain amount of 
dust attenuation to reproduce the observed UV LFs. 
Let us compare the dust attenuation in the mock-UDF12 galaxies with the 
observations to test whether our treatment for the dust attenuation works well 
quantitatively. 

\begin{figure*}
\includegraphics[width = 160mm]{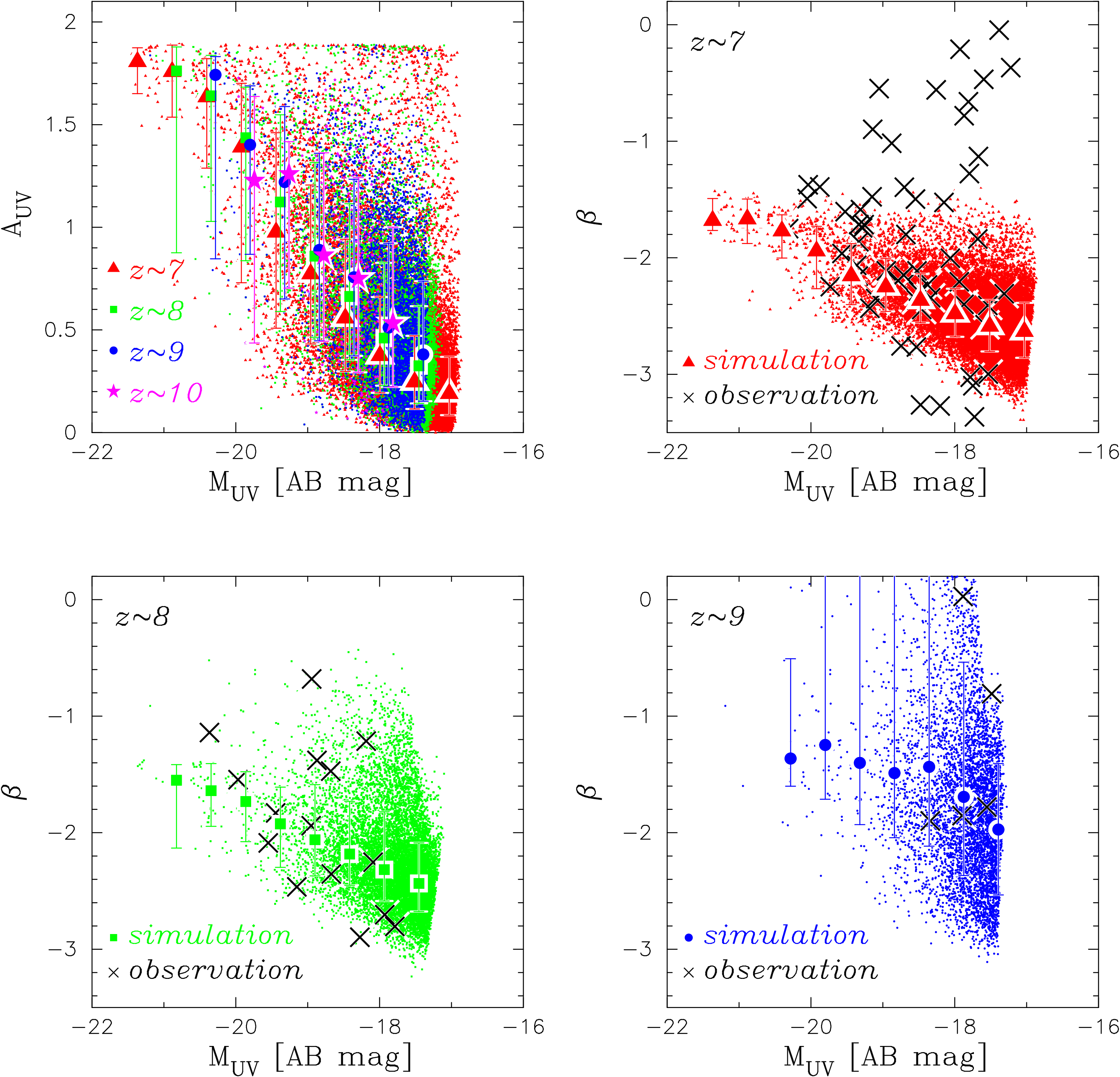}
\caption{The UV dust attenuation (top left panel) and the UV slope $\beta$ 
(all the rest panels) of the mock-UDF12 galaxies  as a function of the absolute 
UV magnitude. The small points are the UV attenuation or the UV slope $\beta$ of each simulated galaxies, 
while the large points with error-bars are the median values and the central $68\%$ ranges in UV magnitude bins.
We also plot the observed UV slope $\beta$ taken from \citet{Dunlop2013} as the cross points. } 
\label{AUV-Beta}
\end{figure*}

The top left panel of Fig. \ref{AUV-Beta} represents the dust attenuation 
$A_{\rm UV}$ at $1500~{\rm \AA}$ as a function of the UV magnitude. 
The small points are the UV attenuation of each simulated galaxy, 
while the large points with error-bars are the median values and the central $68\%$ ranges in UV magnitude bins.  
The range of the dust attenuation is $A_{\rm UV} = 0$ to $A_{\rm UV} = 1.9$. 
This upper limit of $A_{\rm UV}$ comes from our recipe of the UV escape probability 
described in equation (4); we have a lower limit of the probability as 
$(1 - \delta) / 2$ when $\tau_{\rm d} \rightarrow \infty$ and we set $\delta = 0.65$. 
The $A_{\rm UV}$ range corresponds to $A_V = 0$--0.76 under the Calzetti law we assumed 
(i.e., $A_{1500~{\rm \AA}} / A_{5500~{\rm \AA}} = 2.5$). 
Interestingly, the typical value of the dust attenuation is proportional to the UV magnitude. 
We also find that there are no luminous ($M_{\rm UV} \leq -21$) galaxies with less attenuation 
($A_{\rm UV} \leq 1.5$) but faint galaxies are distributed over all the range of $A_{\rm UV}$. 
\cite{Dunlop2013} argue $A_V \sim 0.2$ for the UDF12 galaxies with the 
Calzetti law, which corresponds to $A_{\rm UV} \sim 0.5$.
A similar attenuation is found for $M_{\rm UV} = - 18$ mock-UDF12 galaxies, 
whereas a larger attenuation is found for more UV luminous galaxies. 
Recently, \cite{Taysun2013} argue a possibility of $A_V \sim 1.8$ 
which is more than a factor of 2 larger than our result.
We also find that the dust attenuation is not zero even in galaxies observed at 
$z \sim 10$ in our simulation. 
Note that we have solved the chemical evolution (i.e., dust amount evolution) 
consistently with the star formation activity in galaxies in our simulation 
although we have assumed a constant dust-to-metal ratio which may evolve 
depending on the star formation history \citep{Inoue2011b, Inoue2003}. 

Next we explore the UV slope. 
In order to compare our simulations with the observation, we adopt exactly the same formulae 
as \citet{Dunlop2013} to calculate the UV slope $\beta$: 
\begin{equation}
\beta = 4.43 (J_{125} - H_{160}) - 2, 
\label{eq:betaz78}
\end{equation}
for $z \sim 7$ and $z \sim 8$ galaxies, and 
\begin{equation}
\beta = 9.32 (JH_{140} - H_{160}) - 2, 
\label{eq:betaz9}
\end{equation}
for $z > 8.5$ galaxies. 
The top right, the bottom left and the bottom right panels show the UV slope 
$\beta$ for the mock-UDF12 galaxies at $z \sim 7, ~8$ and $z \sim 9$, respectively. 
The small points are the UV slope $\beta$ of each simulated galaxy, 
while the large points with error-bars are the median values and the central $68\%$ ranges in UV magnitude bins. 
We also plot the observed UV slope $\beta$ (cross points) taken from \cite{Dunlop2013}. 
The uncertainty on $\beta$ for each measurement reaches 1.2 for $z\sim7$ and 
$\sim8$ (eq.~[\ref{eq:betaz78}]) and 2.6 for $z>8.5$ (eq.~[\ref{eq:betaz9}]) at the faintest magnitude. 
At $z \sim 7$, the mock-UDF12 galaxies with $M_{\rm UV} < -20$ have $\beta \simeq -1.6$, 
but less luminous galaxies are distributed over $\beta = -3.4$ to $-1.6$.
This upper limit corresponds to the upper limit of $A_{\rm UV}$ in our recipe.
On the other hand, the observed data are distributed over $\beta = -3.4$ to 0.
Probably this larger scatter is partly because of the observational error, 
while we may underestimate the scatter in the simulated galaxies 
because we employ a  unique dust attenuation model 
in which we adopt the Calzetti extinction law with a single dust geometry for all simulated galaxies. 
The mean value $\beta = -2.1 \pm 0.2$ for $M_{\rm UV} = -18$ reported by \cite{Dunlop2013} 
is consistent with our expectation ($\beta = -2.4 \pm 0.3$) at the same magnitude. 
A similar relation is also found by \citet{Bouwens2013}. 
At $z \sim 8$ and $z \sim 9$, the $\beta$ distribution of the mock-UDF12 galaxies shows a large redward scatter. 
This is because the Ly$\alpha$ break affects the bluer band magnitude in the 
formulae to estimate $\beta$ presented above.
Overall the $\beta$ distribution of the mock-UDF12 galaxies reproduce the observations very well.

\subsection{Star Formation Rate and Specific Star Formation Rate}
\begin{figure*}
\includegraphics[width = 160mm]{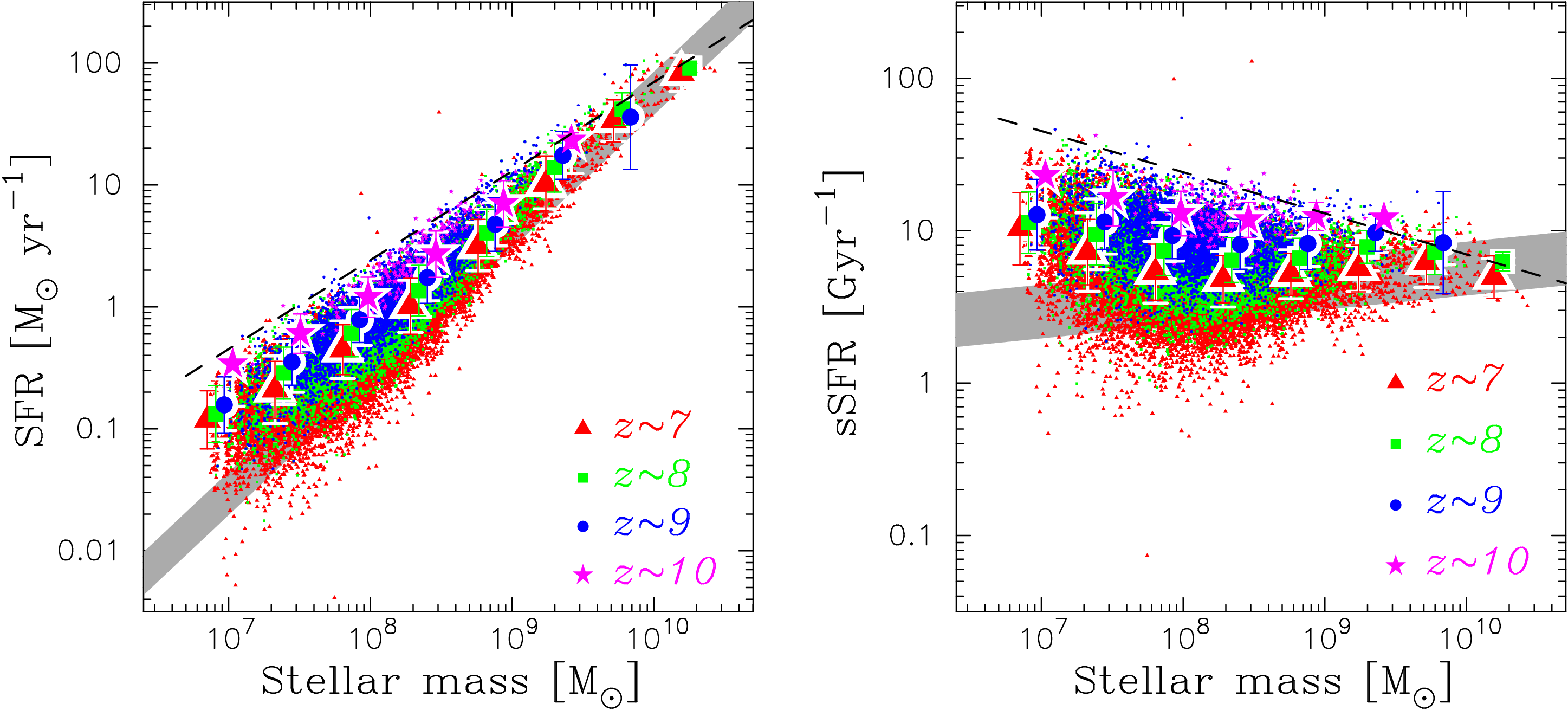}
\caption{The star formation rate (left panel) and the specific star formation rate 
(right panel) as a function of the stellar mass for the mock-UDF12 galaxies. 
The small points are the SFR and sSFR of each simulated galaxy, 
while the large points with error-bars are the median values and the central 68\% ranges in stellar mass bins.
The meanings of the symbols are denoted in the panels and same as in Fig. \ref{UV_HM_SM}. 
The shaded region is our simple models that the star formation rate is determined by 
the halo accretion rate from $z \sim 7$ to $z \sim 10$. 
The dashed lines represent the relations between the SFR or sSFR and the stellar 
mass observed in the LBGs at $z \sim 4$ \citep{Bouwens2012c}. }
\label{SM_SFR_SSFR}
\end{figure*}

The star formation rate (SFR) is very important to study how the stellar components 
in galaxies were build up and how many ionizing photons were produced for the reionization of the Universe. 
Fig. \ref{SM_SFR_SSFR} represents the SFR (left panel) and the specific SFR 
(sSFR; right panel) as a function of the stellar mass for the mock-UDF12 galaxies. 
The small symbols are the same meanings as in Fig. \ref{UV_HM_SM} 
and large symbols with error-bars are the median values and the central 68\% ranges in stellar mass bins. 
The mock-UDF12 galaxies show an almost linear correlation between their SFR 
and stellar mass and form the so-called `main-sequence' for star-forming galaxies 
\citep[e.g.,][]{Brinchmann2004, Noeske2007, Daddi2007}. 
In the left panel of Figure 7, the correlations at different redshifts are 
roughly overlapped each other, while a weak evolution of SFRs from $z \sim 10$ to 
$z \sim 7$ can be seen. This evolution is more clearly seen in the right panel 
showing sSFRs: higher sSFRs as higher-$z$.
This is probably because objects collapsed at higher-$z$ have generally more dense gas, 
thus the SFR becomes higher in such objects. 
On the other hand, the dispersions in sSFRs of the different redshifts are in fact 
similar to each other as found from the 68\% ranges shown. The apparent wider 
distributions for the lower-$z$ samples is just because of a larger number of the galaxies. 

The sSFRs of the mock-UDF12 galaxies are higher than those of `main sequence' 
galaxies at $z < 2$ \citep{Brinchmann2004, Noeske2007, Daddi2007} 
but are similar to those of LBGs at $z \sim 4$ \citep{Bouwens2012c} as shown by 
the dashed lines in Fig. \ref{SM_SFR_SSFR}. 
This indicates that the UDF12 galaxies at $z > 7$ are in a similar starburst 
phase found in LBGs at $z \sim 4$ rather in the normal quiescent phase found at low-$z$.
However, we find that most of our mock-UDF12 galaxies lie in the lower part than the $z\sim4$ observed relations. 

Let us explore the physical process controlling such relations (i.e., `main sequence') for the UDF12 galaxies. 
Suppose that the SFR is proportional to the dark matter (DM) accretion rate estimated by \cite{McBride2009, Ishiyama2013}: 
\begin{equation}
SFR = f_* \left(\frac{\Omega_{\rm B}}{\Omega_{\rm M}} \right) \frac{dM_{\rm halo}}{dt}\,, 
\label{eq:sfr_halo}
\end{equation}
\begin{equation}
\frac{dM_{\rm halo}}{dt} = 1.79 \times 10^{-14} f(z) \left(\frac{M_{\rm halo}}{{\rm M_\odot}} \right)^{1.094},             
\label{eq:dmac}
\end{equation}
and 
\begin{equation}
f(z) = (1 + 1.75z) \sqrt{\Omega_{\rm M}(1 + z)^3 + \Omega_{\rm \Lambda }}, 
\end{equation}
where $f_*$ is the star formation efficiency. 
If we introduce a constant efficiency as $M_* / M_{\rm halo} \sim f_* \Omega_{\rm B} / \Omega_{\rm M} \sim 0.01$, 
then we obtain 
\begin{equation}
SFR(M_*) \sim (4.3_{z \sim 7}, 9.7_{z \sim 10}) \times 10^{-10} \left( \frac{M_*}{{\rm M_\odot}} \right)^{1.094}~{\rm M_{\odot} ~yr^{-1}}\,,
\label{eq:sfr}
\end{equation}
which is shown as the shaded area in the left panel of Fig. \ref{SM_SFR_SSFR}. 
The redshift dependence of eq. (\ref{eq:dmac}) from $z \sim 7$ to $z \sim 10$ is as small as our simulation results. 
We also obtain
\begin{equation}
sSFR(M_*) \sim (0.43_{z \sim 7}, 0.97_{z \sim 10}) \left( \frac{M_*}{{\rm M_\odot}} \right)^{0.094}~{\rm Gyr}^{-1}\,,
\label{eq:ssfr}
\end{equation}
which is the shaded area in the left panel of Fig. \ref{SM_SFR_SSFR}. 
We find a very good agreement between our simulation and the simple estimation 
by equations (\ref{eq:sfr}) and (\ref{eq:ssfr}) at the massive end of the stellar mass, 
but a small difference can be seen. 
This implies that $M_* / M_{\rm halo} \sim f_* \Omega_{\rm B} / \Omega_{\rm M}$ has a weak stellar (halo) mass dependence 
and the baryon physics is also necessary to explain the trend. 

\subsection{Star Formation Rate Density}
\begin{figure}
\includegraphics[width = 80mm]{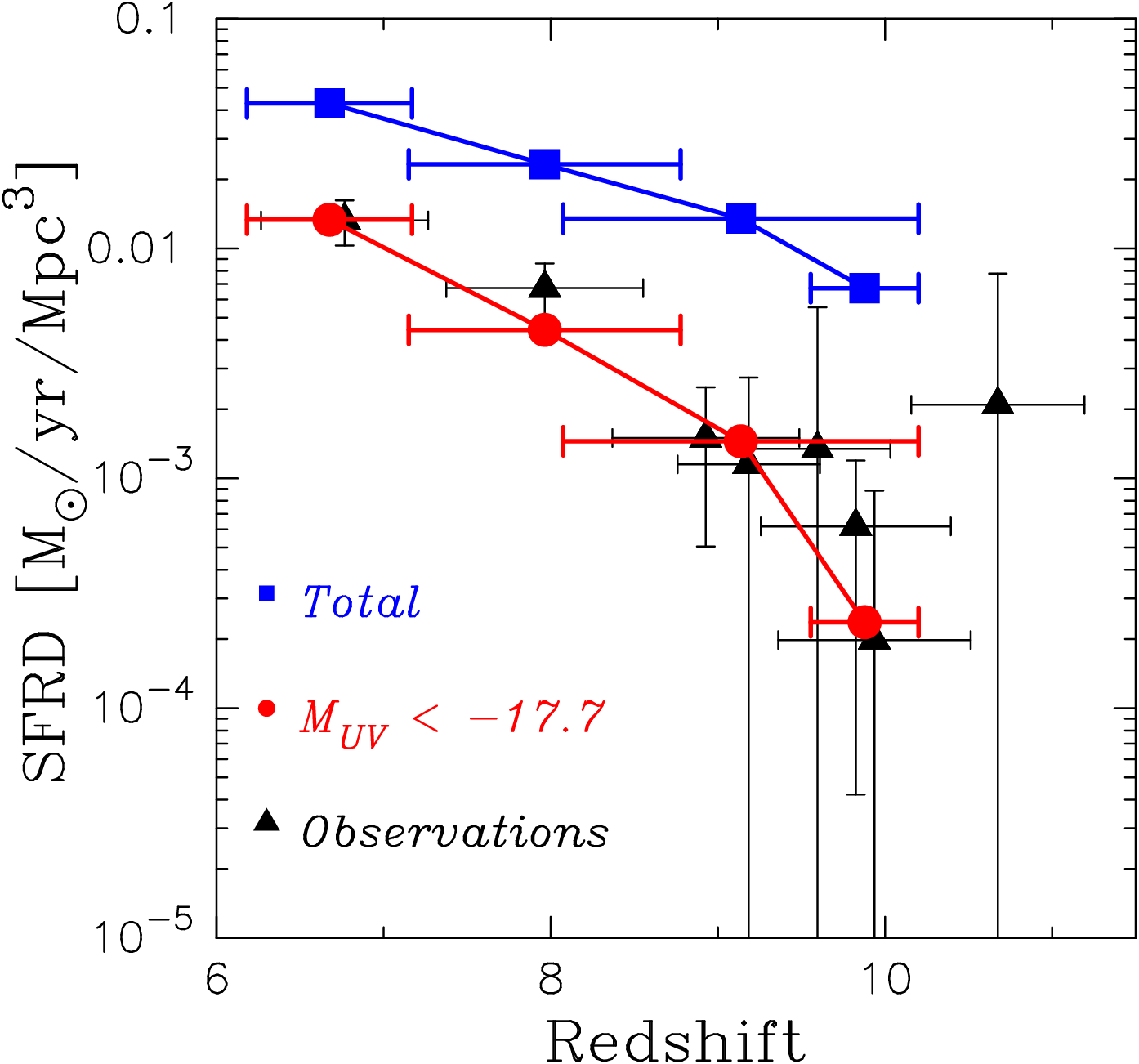}
\caption{The evolution of the star formation rate density (SFRD) from $z \sim 10$ 
to $z \sim 7$. 
The circle and square points are the SFRD integrated over $M_{\rm UV} < -17.7$ 
and the total SFRD in our simulation, respectively. 
The horizontal error-bars represent the redshift coverage of the corresponding 
color selection criteria.
The triangles represent observational estimates with $M_{\rm UV} < -17.7$ 
\citep{Bouwens2007, Bouwens2012a, Bouwens2012b, Coe2013, Ellis2013, Zheng2012, Oesch2013}. }  
\label{SFRD}
\end{figure}

The star formation rate density (SFRD) is also useful to check the calibration 
of our simulation as well as the stellar mass function and the UV LFs. 
Fig. \ref{SFRD} shows the evolution of the SFRD from $z \sim 10$ to $z \sim 7$. 
The triangles represent observational estimates based on galaxies with 
$M_{\rm UV} < -17.7$ 
\citep{Bouwens2007, Bouwens2012a, Bouwens2012b, Coe2013, Ellis2013, Zheng2012, Oesch2013}. 
The all observational data are corrected for dust attenuation. 
Thus, the simulated data do not include any dust attenuation. 
The circle and square points are the SFRD of the galaxies with $M_{\rm UV} < -17.7$ 
as the observations and the total SFRD in our simulation, respectively. 
Our results for $M_{\rm UV} < -17.7$ reasonably agree with the observations within 
the errors, suggesting that our calibration is good enough.
One important finding from the comparison is that the SFRD of the galaxies 
with $M_{\rm UV} < -17.7$ is only a minor part of the total SFRD even at $z \sim 7$. 
Therefore, we still miss a major part of the SFRD even in the current deepest survey.

\subsection{Mass Weighted Age}
\begin{figure}
\includegraphics[width = 80mm]{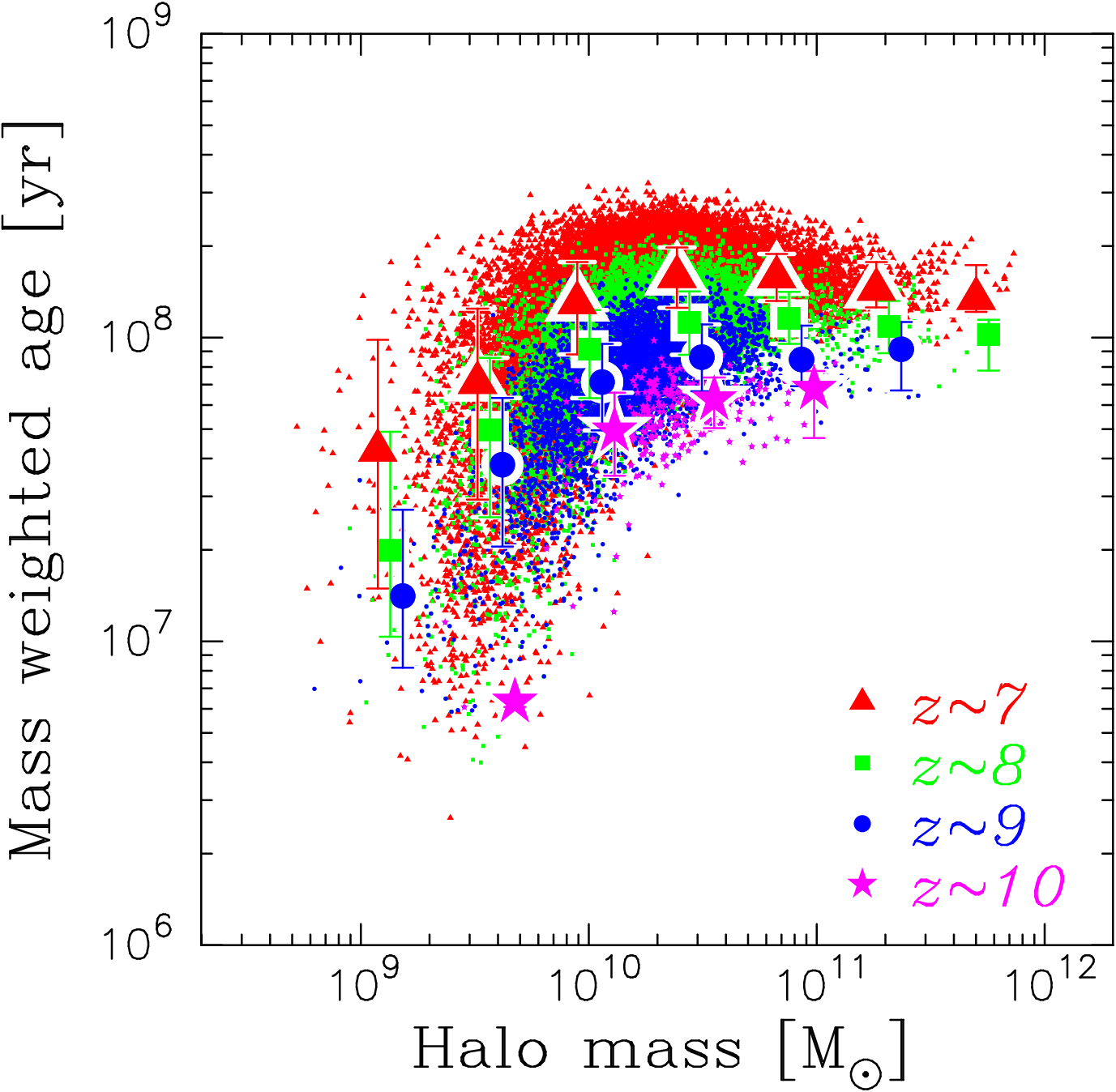}
\caption{The mass weighted age of the mock-UDF12 galaxies as a function of 
their host halo mass. The symbols are the same as in Fig. \ref{UV_HM_SM}.} 
\label{HM_AGE}
\end{figure}

We compute the mass weighted age $\tau_{\rm age}$ of each mock-UDF12 galaxy to characterize the stellar population. 
The definition of the age is given by the following equation: 
\begin{equation}
\tau_{\rm age} = \frac{\sum_i t_i^{\rm age} m_i^{\rm star}}{\sum_i m_i^{\rm star}}, 
\label{eq:age}
\end{equation}
where $t_i^{\rm age}$ and $m_i^{\rm star}$ are the age and mass of the {\it i}th 
star cluster in a simulated galaxy. 
Fig. \ref{HM_AGE} shows the mass weighted age of the mock-UDF12 galaxies 
as a function of their halo mass. 
The symbols are the same as Fig. \ref{UV_HM_SM}. 
The typical ages of the mock-UDF12 galaxies are $180 ~{\rm Myr} ~(z \sim 7)$, 
$100 ~{\rm Myr}~(z \sim 8)$, $80 ~{\rm Myr}~(z \sim 9)$ and $50 ~{\rm Myr}~(z \sim 10)$, respectively. 
The ages decrease with increasing redshift. 
On the other hand, there is a large dispersion found at halo masses less than about $10^{10}~M_\odot$. 
This is because the SN feedback effectively works in such smaller mass galaxies 
and then the star formation occurs stochastically in such galaxies 
(see also the next subsection). 
Similar discussion can be seen in \citet{Jaacks2012b}. 
In addition, the mass resolution of our simulation may not be enough to resolve 
such smaller mass galaxies and may enlarge the dispersion.

\subsection{Star Formation History}
\begin{figure*}
\includegraphics[width = 160mm]{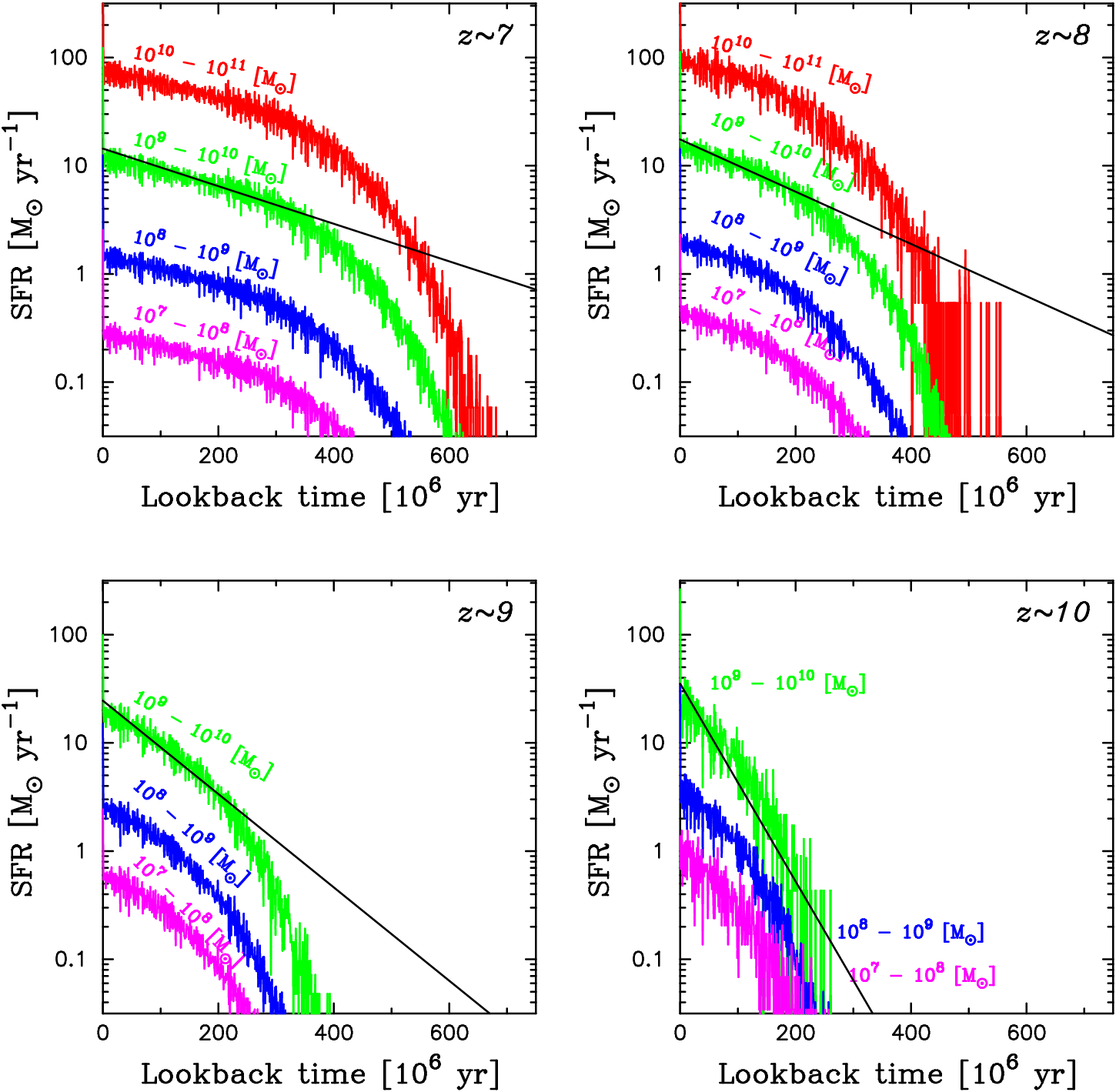}
\caption{The average star formation history of the mock-UDF12 galaxies 
within a certain mass range. 
The lines in each panel at $z \sim 7$ and $z \sim 8$ show the SHFs of simulated galaxies 
within the stellar mass range of $10^{10}$--$10^{11}~{\rm M_{\odot}}$, $10^{9}$--$10^{10}~{\rm M_{\odot}}$ 
$10^{8}$--$10^{9}~{\rm M_{\odot}}$ and $10^{7}$--$10^{8}~{\rm M_{\odot}}$ from the top to the bottom. 
The lines at $z \sim 9$ and $z \sim 10$ show the SFHs within the range of $10^{9}$--$10^{10}~{\rm M_{\odot}}$ 
$10^{8}$--$10^{9}~{\rm M_{\odot}}$ and $10^{7}$--$10^{8}~{\rm M_{\odot}}$ from the top to the bottom. 
The time bin of this calculation is $10^6~{\rm yr}$. The solid lines of each panel 
is an exponential fitting.}
\label{SM_SFH}
\end{figure*}

One has to assume a star formation history to perform the SED fitting method 
which is often used to explore the physical properties such as the stellar mass 
and the SFR of high-$z$ galaxies. 
A constant SFR or exponentially decreasing SFR are usually assumed.
However, recent studies suggest an increasing SFR with time for high-$z$ galaxies 
\citep[e.g.,][]{Reddy2012}.
Thus, it is very important to examine the SFH from theoretical point of view. 
Fig. \ref{SM_SFH} represents average SFHs of the mock-UDF12 galaxies for a 
certain mass range and a redshift. 
The lines in each panel at $z \sim 7$ and $z \sim 8$ show the SFHs of simulated galaxies 
within the stellar mass range of $10^{10}$--$10^{11}~{\rm M_{\odot}}$, $10^{9}$--$10^{10}~{\rm M_{\odot}}$,  
$10^{8}$--$10^{9}~{\rm M_{\odot}}$ and $10^{7}$--$10^{8}~{\rm M_{\odot}}$ from the top to the bottom. 
The lines at $z \sim 9$ and $z \sim 10$ show the SFHs within the range of $10^{9}$--$10^{10}~{\rm M_{\odot}}$, 
$10^{8}$--$10^{9}~{\rm M_{\odot}}$ and $10^{7}$--$10^{8}~{\rm M_{\odot}}$ from the top to the bottom. 
The time resolution of the SFHs is set to be $10^6$ yr to see the stochastic star formation activity. 
As found in Fig. \ref{SM_SFH}, the SFR clearly increases with time. 
This suggests that it is strongly preferable to choose increasing SFR with the time 
(e.g., exponentially increasing) for the SED fitting method. 
This result is consistent with previous theoretical work \citep{Jaacks2012b, Dayal2013}.
In order to examine the typical time-scale of the star formation, we adopt 
an exponential function:  
\begin{equation}
SFR(t) = a \exp{\left( \frac{t}{\tau_{\rm SF}} \right)},
\label{eq:sfh} 
\end{equation}
where $a$ and $\tau_{\rm SF}$ are the normalization parameter and the typical 
star formation time-scale, respectively. 
We obtain these values with a $\chi^2$ method. 
The typical star formation time-scales $\tau_{\rm SF}$ obtained are 
$250~ {\rm Myr}$ ($z \sim 7$), $180 ~{\rm Myr}$ ($z \sim 8$), $100 ~{\rm Myr}$ 
($z \sim 9$) and $50 ~{\rm Myr}$ ($z \sim 10$). 
These values are almost constant among different mass ranges if the redshift is the same 
and are somewhat longer than but still similar to the mass weighted age obtained 
in the previous subsection. 

\subsection{Metallicity}
Observationally, the metallicity of galaxies is determined by measuring a ratio 
(or ratios) of emission or absorption lines.
Emission lines are usually emit from the HII regions and the photo-dissociation 
regions (PDRs). 
Thus, these lines come from the star forming region which is relatively dense 
region in a galaxy. 
In order to compare such observations with our model properly, 
we need to calculate the metallicity in star forming regions in a simulated galaxy 
rather than that in the whole of the galaxy.
We here introduce a new definition of the metallicity for simulated galaxies 
as following equation: 
\begin{equation}
Z_{\rm neb} = \frac{\sum_{i} Z_{i} L_{i}^{\rm LyC}}{\sum_{i} L_{i}^{\rm LyC}}, 
\label{eq:metal}
\end{equation}
where $Z_{i}$ and $L_{i}^{\rm LyC}$ are the metallicity and the intrinsic 
Lyman continuum luminosity of $i$th star particle in a simulated galaxy. 
We have assumed that the gas metallicity around a star particle is the same as 
that of the star particle.
In this definition, only the metal around the very young star particles is 
taken into account and the metal around aged star particles and the metal far 
from star forming regions do not contribute to the metallicity. 
The metallicity of this definition is usually higher than that averaged over 
the whole of a simulated galaxy. 
We call this metallicity the nebular metallicity.

\begin{figure}
\includegraphics[width = 80mm]{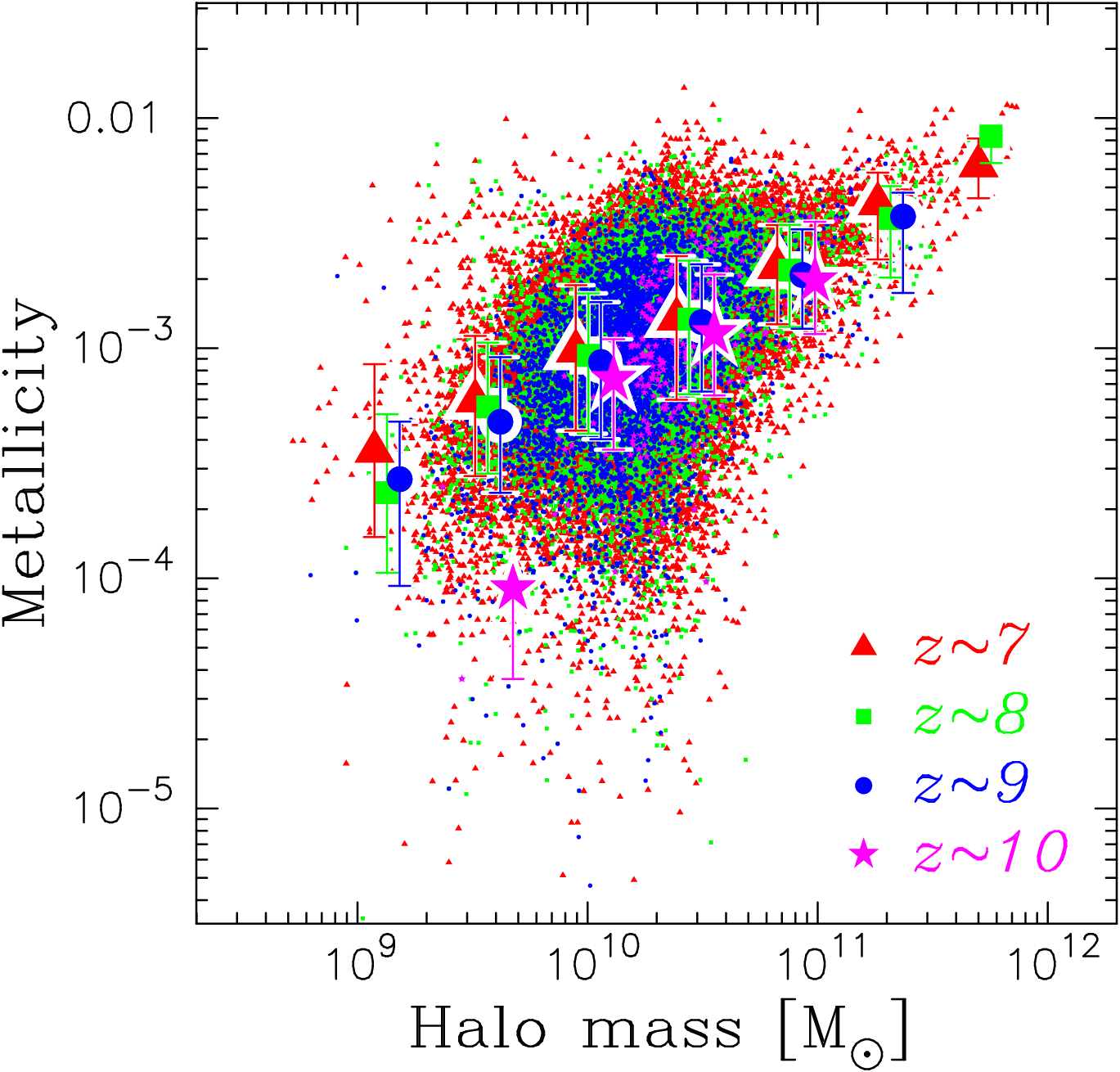}
\caption{The nebular metallicity of the mock-UDF12 galaxies as a function of 
their halo mass. The symbols are the same as Fig. \ref{UV_HM_SM}.}
\label{HM_METAL}
\end{figure}

Fig. \ref{HM_METAL} represents the nebular metallicity of the mock-UDF12 
galaxies as a function of their host halo mass. 
The symbols are the same as Fig. \ref{UV_HM_SM}.
The nebular metallicity is roughly proportional to their host halo mass 
although the dispersion is large. 
We also find that the metallicity evolution from $z \sim 10$ to $z \sim 7$ is weak as with the SFR. 
Interestingly, the metallicity of some galaxies reaches at $> 0.1$ solar 
metallicity ($Z_\odot = 0.02$) even at $z \sim 10$. 
These suggest that the metal enrichment is a rapid process. 

\subsection{Morphology and Size}
\begin{figure*}
\includegraphics[width = 160mm]{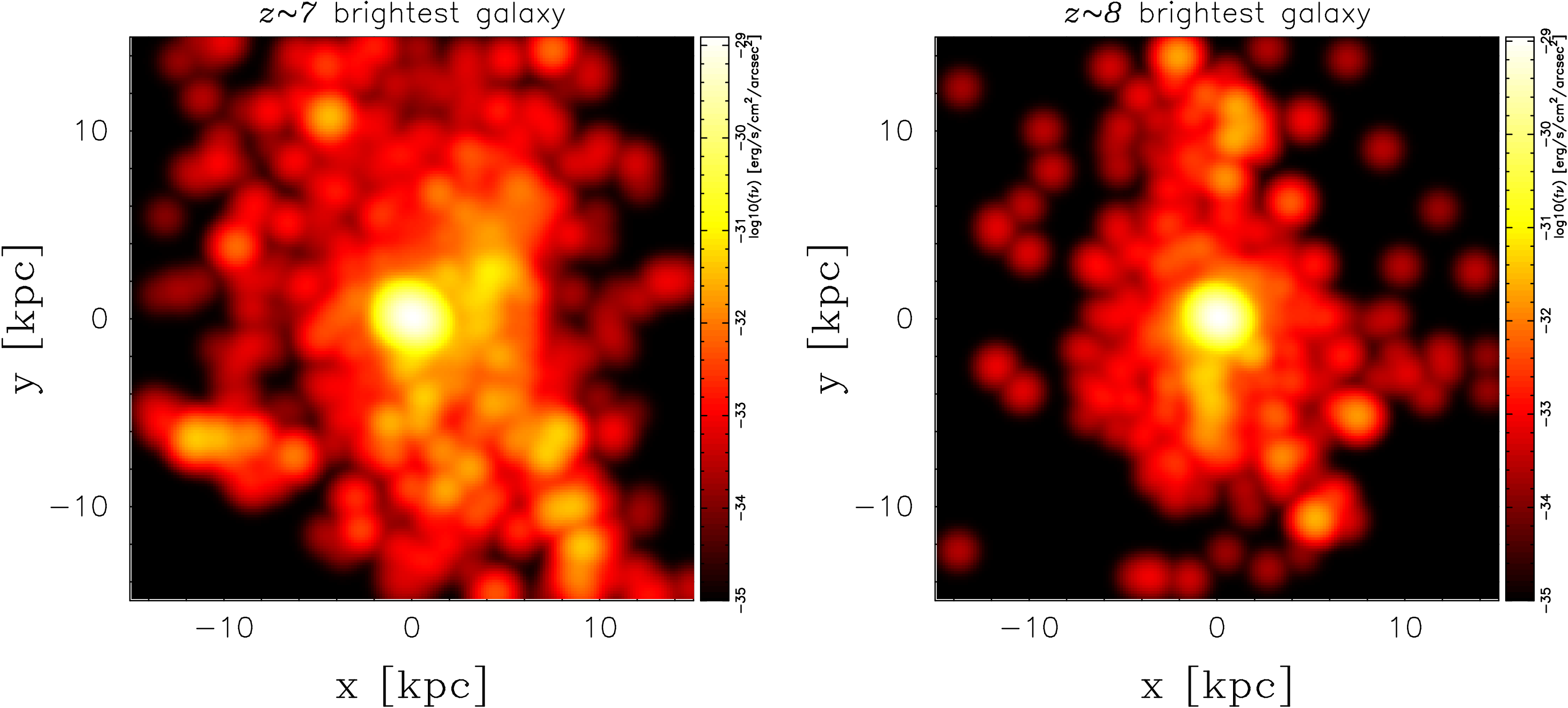}
\caption{The surface brightness distributions 
at the rest-frame $1500~{\rm \AA}$ of the UV brightest galaxy among the mock-UDF12 galaxies 
at $z \sim 7$ (the left panel) and $z \sim 8$ (the right panel), respectively. }  
\label{MORPHOLOGY}
\end{figure*}

\begin{figure*}
\includegraphics[width = 160mm]{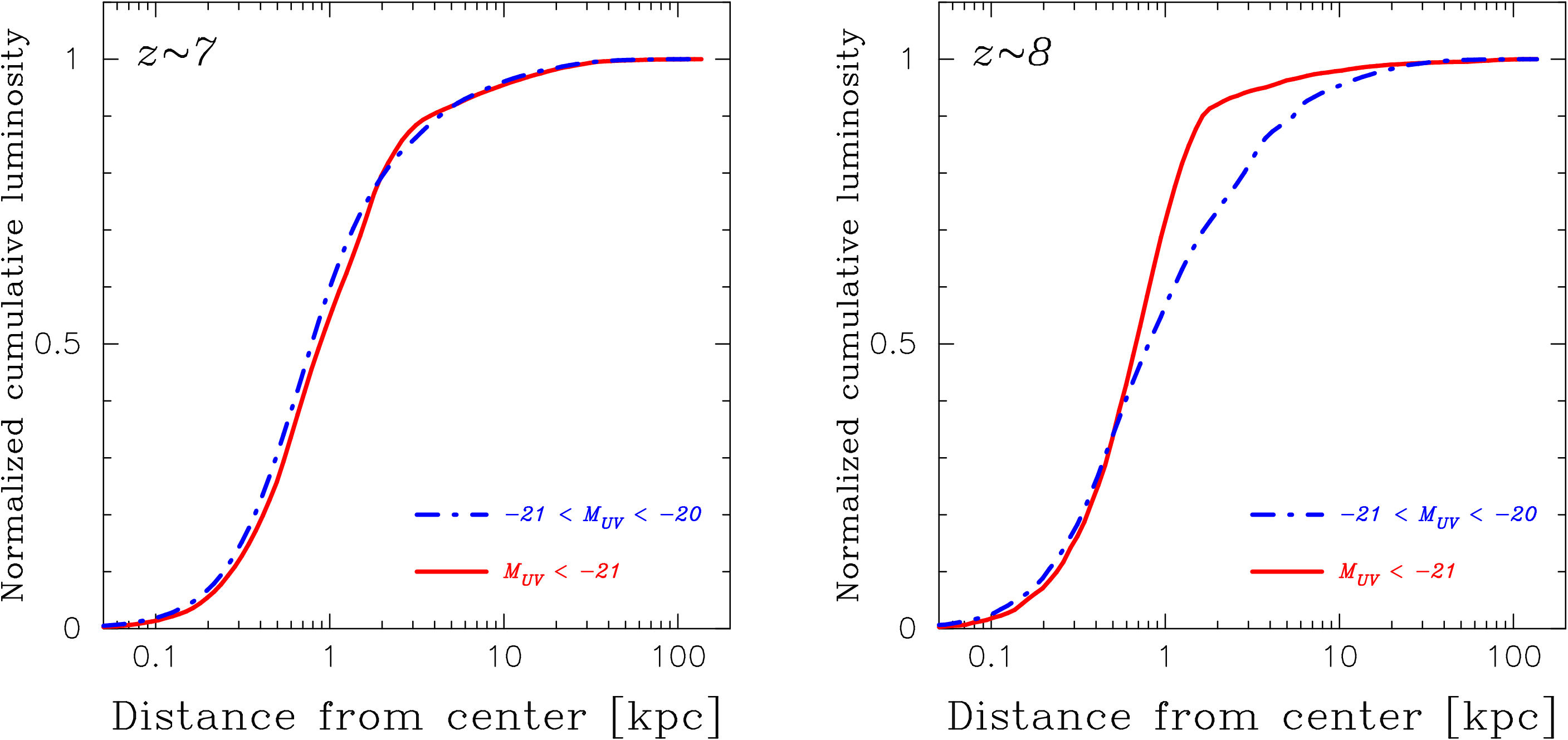}
\caption{The azimuthally summed up and normalized cumulative profiles with two $M_{\rm UV}$ ranges 
at $z \sim 7$ (the left panel) and $z \sim 8$ (the right panel).  
The solid and dashed lines are the profiles of the stacked 
images with $M_{\rm UV} < -21$ and $-21 < M_{\rm UV} < -20$, respectively. }  
\label{SIZE}
\end{figure*}

In order to explore  the size and the morphology of the mock-UDF12 galaxies, 
it is necessary to resolve the inner structure of the simulated galaxies. 
Thus, we discuss only massive (UV bright: $M_{\rm UV} < -20$) galaxies which are 
composed of at least about $1,000$ star particles. 
Fig. \ref{MORPHOLOGY} shows the surface brightness distributions 
at the rest-frame $1500~{\rm \AA}$ of the UV brightest galaxy among the mock-UDF12 galaxies 
at $z \sim 7$ (the left panel) and $z \sim 8$ (the right panel), respectively. 
We have applied a Gaussian smoothing with $\sigma_{\rm G} = 0.1"$ in order to 
take the point-spread-function of the HST/WFC3 approximately into account. 
The images of the UV brightest mock-UDF12 galaxies are roughly round 
but still show somewhat clumpy and elongated structure, in particular in the $z \sim 7$ case.
Interestingly, such a shape has been found in the observational data \citep{Ono2013}. 
To compare with observational studies (the brightness profile), 
we make a stacked UV images of the mock-UDF12 galaxies.
To do so, we stack the distributions of star particles projected 
on the sky using the light-cone output of the mock-UDF12 galaxies.
In the stacking process, we regard the luminosity center of each galaxy as 
the galaxy center and we do not choose any specific direction of the galaxy 
(i.e., the position angles in the stacking are random).
The spatial unit of the galaxies is the physical unit not comoving unit. 
In our simulation box, the numbers of galaxies with $M_{\rm UV} < -21$ and 
$-21 < M_{\rm UV} < -20$ are $32~(6)$ and $155~(45)$ at $z \sim 7~(z \sim 8)$, respectively. 
Fig. \ref{SIZE} shows the azimuthally summed up and 
normalized cumulative profiles of the stacked images of the simulated galaxies 
at $z \sim 7$ (left panel) and $z \sim 8$ (right panel). 
The solid and dashed lines represent the profiles for $M_{\rm UV} < -21$ and 
$-21 < M_{\rm UV} < -20$, respectively. 
The half light radii of the stacked images with $M_{\rm UV} < -21$ 
$(-21 < M_{\rm UV} < -20)$ are $0.60~{\rm kpc}$ $(0.55~{\rm kpc})$ at $z \sim 7$ 
and $0.48~{\rm kpc}$ $(0.55~{\rm kpc})$ at $z \sim 8$, respectively. 
These results are also similar to the results of the UDF12 campaign \citep{Ono2013}. 

\section{Discussion}
We have presented some physical properties of the UDF12 galaxies at $z \sim 7$--10 
expected from our cosmological hydrodynamic simulation.
Here we discuss a few additional issues derived from the simulation.

\subsection{$z > 10$ Galaxies}
So far, we have studied only upto $z \sim 10$ selected galaxies. 
\citet{Oesch2013} \citep[see also][]{Bouwens2011,Ellis2013} presented one 
$z = 11.8$ galaxy candidate (XDFjh-39546284). 
Thus, it is also interesting to study the physical properties of such 
$z > 10$ galaxies not only for HST deep survey but also for future observations 
with the James Webb Space Telescope and $30~ \rm m$ telescopes. 
Here, we use the same color selection criterion as \citet{Oesch2013}: 
\begin{equation}
JH_{140} - H_{160} > 1.0.
\end{equation}
Moreover, we adopt the detection limit criterion of $H_{\rm 160} < 30$. 
As a result, we find $81$ simulated galaxies from $z = 10.6$ to $z = 11.7$ 
in our simulation box. 
All their $H_{160}$ band magnitudes are fainter than $28.6$. 
We note that our field-of-view (FOV) is about $400~ {\rm arcmin^2}$ at the redshift 
and then the expected number is about two galaxies per one FOV of the UDF12 survey. 
Therefore, the detection of one such highest-$z$ galaxy in the UDF12 campaign 
is consistent with our simulation if we consider the Poisson error. 
The halo and stellar masses of these galaxies are 
$5 \times 10^{9} - 1 \times 10^{11}~ {\rm M_{\odot}}$ and 
$2 \times 10^7 - 2 \times 10^9~ {\rm M_{\odot}}$, respectively. 
The stellar mass weighted age and the nebular metallicity $Z_{\rm neb}$ are $1 \times 10^7 - 8 \times 10^7 ~{\rm yr}$ 
and $0.00005 - 0.003$, respectively. 
Clearly, $z > 10$ galaxies selected by the drop-out technique are very early evolutionary phase objects.

\subsection{Prediction for ALMA}
As discussed in section 4.2, the mock-UDF12 galaxies suffer a modest amount 
of dust attenuation.
Therefore, these galaxies also emit the thermal radiation from dust in the far-infrared wavelength. 
They could be detectable with the Atacama Large Milimetre-submilimetre Array (ALMA).
Submm observations of very high-$z$ galaxies is also important 
to know the chemical enrichment and the dust formation at very high-$z$. 
According to \citet{Shimizu2012}, the submm flux of star-forming galaxies 
with ${\rm SFR} \sim 100~{\rm M_{\odot}/yr}$ is about $1~{\rm mJy}$. 
Indeed, some of our simulated galaxies have ${\rm SFR} \sim 100~{\rm M_{\odot}/yr}$ 
and the galaxies would be detectable with ALMA. 
We calculate the submm flux of the simulated galaxies by the same procedure as 
\citet{Shimizu2012}. 

\begin{figure}
\includegraphics[width = 80mm]{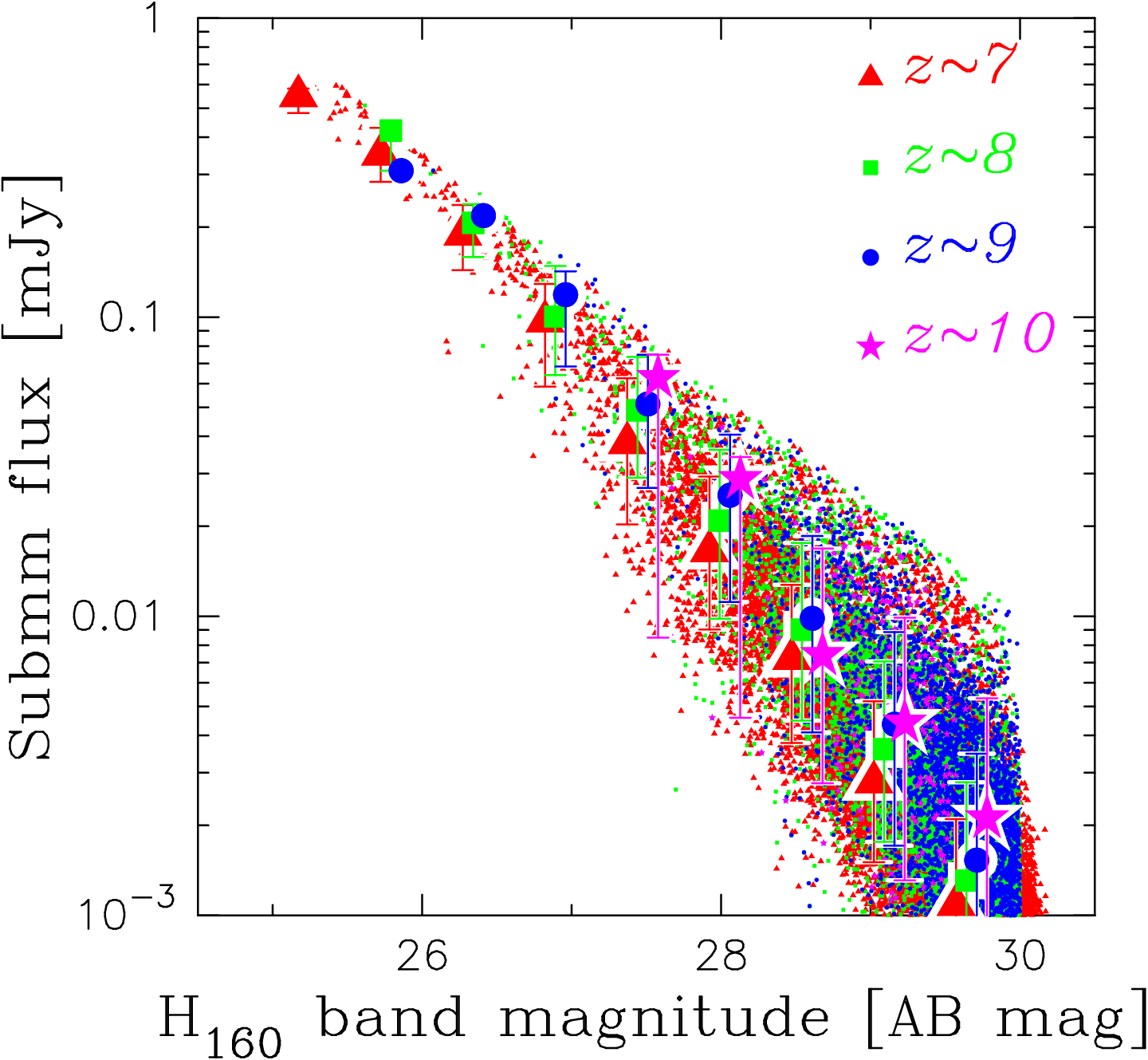}
\caption{The expected submm flux density at $350~{\rm GHz}$ in the observer's 
frame of the mock-UDF12 galaxies as a function of the observed $H_{160}$ band magnitude. 
The symbols are the same as Fig. \ref{UV_HM_SM}.}
\label{SUBMM}
\end{figure}

The dust temperature for our sample galaxies is typically $\sim 30~ \rm K$.  
Fig. \ref{SUBMM} shows the expected submm flux density at $350~{\rm GHz}$ 
in the observer's frame as a function of the observed $H_{160}$ band magnitude. 
We find a tight correlation between the two quantities independent of the redshift. 
If we set the detection limit as 0.1 mJy, galaxies brighter than $H_{160} = 27.5$ may be detected. 
However, most of the galaxies with $H_{160} = 27.5$ actually have much fainter submm flux density. 
In order to increase the probability of the detection, 
we should observe galaxies with $H_{160} > 27$ or even 26.5. 
Actually, the brightest $H_{160}$ magnitude in the UDF12 survey is $H_{160} = 27.2$. 
However, other wide-field surveys like the Brightest of Reionizing Galaxies (BoRG) 
survey \citep{Trenti2011, Trenti2012, Bradley2012} and surveys for 
gravitationally-lensed galaxies like the Cluster Lensing And Supernova survey with 
Hubble (CLASH) survey \citep{Postman2012} have reported detection of some galaxies with 
$H_{160} < 27$ at $z > 7$. 
These galaxies would be good targets for ALMA. 

\subsection{Galaxies selected and not selected by the UDF12 Survey}
So far we have discussed the galaxies detected and color-selected by the UDF12 campaign.
As shown in Fig. \ref{SEL_EFFICIENCY}, however, most of the simulated galaxies 
are not selected owing to either detection limit or the color criteria. 
Indeed, the number fraction of the selected galaxies is about $12 \%$ for $z \sim 7$ 
and about $2 \%$ for $z \sim 10$. 
What about the corresponding fraction in terms of the cosmic star formation density (SFRD)? 
Fig. \ref{SFRD} shows the evolution of the SFRD from $z \sim 10$ to $z \sim 7$. 
There, we consider an additional condition of $M_{\rm UV} < -17.7$  
in order to directly compare our simulation with the observations. 
This condition is about one magnitude brighter than the detection limit of the 
UDF12 campaign. 
In this subsection, we measure the contribution of the UDF12 galaxies 
to the cosmic SFRD quantitatively. 
We find that the contributions to the total SFRD of the mock-UDF12 galaxies are $52\%, ~39\%, ~30\%$ and $12\%$ 
for $z \sim 7, ~8, ~9$ and $z \sim 10$ , respectively. 
This means that we do not see faint galaxies which contributed to the dominant part of the SFRD. 
Note that we can resolve only halos with $> 10^{9}~{\rm M_{\odot}}$ in our simulation. 
Thus, the relative contribution of the detectable galaxies could be even smaller 
if there are a significant amount of less massive galaxies than our simulation. 

Let us study the properties of galaxies unselected by the UDF12 survey. 
There are four cases: (1) galaxies satisfy the color selection 
criteria and are bright enough to be detected, (2) galaxies satisfy the color 
selection criteria but are too faint to be detected, (3) galaxies do not satisfy 
the color criteria but are bright enough to be detected and (4) galaxies do not 
satisfy neither color criteria nor the brightness limit. 
Table \ref{FRACTION_TABLE} represents the contributions of each case to the SFRD.
We find that the simulated galaxies satisfying the color criteria (cases [1] and [2]) 
contribute to a major fraction of the SFRD (85--95\%) but most of it comes from 
galaxies fainter than the current detection limit (i.e., case [2]), 
except at $z = 7$ where the detected and not-detected galaxies have similar contributions. 
Many of these faint galaxies in the case (2) will be detected by the next 
generation telescopes.
On the other hand, the simulated galaxies of the cases (3) and (4) contribute to 
only a minor fraction of the SFRD (5--15\%). 
Interestingly, all these galaxies reside around the redshift boundaries defined by 
the color selection and there are no galaxies around middle of the redshift ranges. 
The deviation from color selection criteria of the cases (3) and (4) galaxies is less than 0.4 mag. 
Thus, if the selection criteria are somewhat relaxed, 
the galaxies bright enough (i.e., case [3]) can be selected by the color selection. 
In other words, there are no galaxies with very red UV colors at $z \sim 7, ~8, ~9$ 
and $z \sim 10$ in our simulation. 

Finally, we provide a prediction after the James Webb Space Telescope (JWST) starts operation. 
The values in parentheses of Table \ref{FRACTION_TABLE} represent 
the star formation contribution of each case with the detection limit of $< 32$ which corresponds to the JWST one. 
We find that the JWST can detect more than half (72--77\%) of star formation activity. 
Yet, much fainter galaxies accounting for $\sim 28\%$ of the $z \sim 10$ star forming activity will still remain undetectable even with the JWST. 

\begin{table*}
\begin{tabular}{|l|c|r|r||r|r|} \hline
fraction & redshift & case (1) & case (2) & case (3) & case (4) \\ \hline \hline
 & $z \sim 7$ & 0.517 (0.774) & 0.347 (0.089) & 0.086 (0.122) & 0.051 (0.015) \\ \cline{2-6}
$P_{\rm SFRD}$ & $z \sim 8$ & 0.388 (0.749) & 0.460 (0.098) & 0.071 (0.138) & 0.081 (0.015) \\ \cline{2-6}
 & $z \sim 9$ & 0.295 (0.793) & 0.646 (0.149) & 0.019 (0.055) & 0.039 (0.003) \\ \cline{2-6}
 & $z \sim 10$ & 0.122 (0.723) & 0.825 (0.224) & 0.008 (0.047) & 0.045 (0.006) \\ \hline \hline 
\end{tabular}
\caption{The fraction in the star formation rate density ($P_{\rm SFRD}$) of the following four cases: 
(1) galaxies satisfy the color selection criteria and are bright enough to be 
detected, (2) galaxies satisfy the color selection criteria but are too faint 
to be detected, (3) galaxies do not satisfy the color criteria but are bright 
enough to be detected, and (4) galaxies do not satisfy neither color criteria 
nor the brightness limit. 
The values without and with parentheses are the fractions for the detection limits of $< 30$ and $< 32$, respectively. 
}
\label{FRACTION_TABLE}
\end{table*}

\section{Conclusion}
We have made a cosmological hydrodynamic simulation tailored to examine the 
physical properties of very high-$z$ galaxies detected by the deep survey with 
HST/WFC3 performed in 2012, the so-called UDF12 campaign 
\citep{Ellis2013, McLure2013, Schenker2013, Dunlop2013, Robertson2013, Ono2013, Koekemoer2013, Oesch2013}.
Our simulation code follows the formation and the evolution of star forming 
galaxies by new feedback models based on \citet{Okamoto2010} and can reproduce 
the statistical properties of star forming galaxies such as Lyman break galaxies 
(LBGs), Lyman $\alpha$ emitters and sub-mm galaxies \citep{Shimizu2011,Shimizu2012}.
First, we calibrate our model parameters so as to reproduce not only the stellar 
mass function at $z \sim 7$ but also the UV luminosity function from $z \sim 7$ 
to $z \sim 10$. 
Then, we generate a light-cone output which extends from $z = 6$ to $z = 12$ 
using a number of simulation outputs in order to directly compare our model 
with the observations.
It is the most important point of this study that we adopt the same 
color selection criteria as the UDF12 campaign to select galaxies from our 
simulation.
We call the selected galaxies the mock-UDF12 galaxies.
Using these simulated galaxies as a proxy of the observed UDF12 galaxies, 
we can examine the physical properties such as the halo and stellar masses, 
the specific star formation rate, the star formation history, etc. of the real 
galaxies.
This is a new approach to reveal the physical properties of the highest-$z$ galaxies 
for which the so-called SED fitting method may not be reliable because we have 
few photometric data points only in the rest-frame UV.

We find that the observed UV luminosity of the mock-UDF12 galaxies 
is almost linearly proportional to their host halo mass and the correlation does 
not change so much along the redshift from $z \sim 7$ to $z \sim 10$.
For example, the halo mass at $M_{\rm UV} = -21$ is about $4 \times 10^{11}~M_\odot$.
The stellar mass of the galaxies is also proportional to the UV luminosity but 
the power-law slope is about 1.4.
We have derived the fitting formulae for the halo and stellar masses as a function 
of the UV absolute magnitude not corrected for the dust attenuation.
We also find that many mock-UDF12 galaxies are affected by the dust attenuation. 
The amount of the attenuation well correlates with the UV magnitude but shows a large dispersion ranging from 
$A_{\rm UV} = 0$ to 1.9 (or $A_V = 0$ to 0.76) which is larger than that argued by 
\cite{Dunlop2013}.
Nevertheless, our simulation reproduces the observed UV slope $\beta$ distribution 
reported by \cite{Dunlop2013}. 
The mean UV slope also correlates with the UV magnitude. 
Such a trend consistent with the observational results \citep{Bouwens2013}. 
There is no luminous ($M_{\rm UV} < -21$) and less attenuation ($A_{\rm UV} < 1.5$) galaxies in our simulation. 

The star formation rate (SFR) is proportional to the stellar mass with a power-law slope about unity. 
The specific star formation rate (sSFR) shows a very weak dependency on the stellar 
mass and the value is about $10~{\rm Gyr^{-1}}$ which is higher than that of 
the main-sequence for star-forming galaxies at $z<2$ 
\citep[e.g.,][]{Brinchmann2004,Noeske2007,Daddi2007} but is similar to that 
of LBGs at $z > 4$ \citep[e.g.,][]{Bouwens2012c}. 
This indicates that LBGs from $z \sim 4$ to $z \sim 10$ are in a similar starburst phase. 
With the hypothesis of the SFR proportional to the dark matter accretion rate, 
these relations between the stellar mass and the SFR or the sSFR are reproduced reasonably well.
This supports the hypothesis. 
However, the slight gap between our model and the simple model means that the baryon physics 
is also necessary to reproduce the trend. 
The evolution of the cosmic SFR density from $z \sim 7$ to $z \sim 10$ 
in our simulation reasonably matches with the recent observed results 
\citep{Bouwens2007, Bouwens2012a, Bouwens2012b, Coe2013, Ellis2013, Zheng2012, Oesch2013}. 

We have also examined the star formation history (SFH) of the mock-UDF12 galaxies 
and found an increasing SFR with time.
This trend is independent of the redshift and the mass.
Such an increasing SFH is also suggested by recent observational studies 
\citep[e.g.,][]{Reddy2012}.
Therefore we conclude that an increasing SFH should be assumed in the SED fitting 
work where usually a constant or decreasing SFH has been assumed so far.
If we fit the increasing SFH by an exponential function, the typical time-scales 
for $z \sim 7, ~8, ~9$ and $z \sim 10$ galaxies are $250~ {\rm Myr}$, 
$180 ~{\rm Myr}$, $100 ~{\rm Myr}$ and $50 ~{\rm Myr}$, respectively.
We also find that the mass-weighted average ages of the mock-UDF12 galaxies 
are $180 ~{\rm Myr} ~(z \sim 7)$, $100 ~{\rm Myr}~(z \sim 8)$, 
$80 ~{\rm Myr}~(z \sim 9)$ and $50 ~{\rm Myr}~(z \sim 10)$. 
These time-scales are roughly consistent with each other. 

We have introduced an average metallicity weighted by Lyman continuum 
luminosity. 
We call it the nebular metallicity which is the one usually measured 
from emission line ratios observationally.
Interestingly, the nebular metallicity of some mock-UDF12 galaxies is $0.1$ to 
$0.5$ solar metallicity. 
The result implies that the metal enrichment proceeds rapidly even at $z \sim 10$. 
We have also explored the size and morphology of the mock-UDF12 galaxies. 
We find that the UV brightest galaxies at $z \sim 7$ and $z\sim 8$ 
are roughly round but still show clumpiness and asymmetry. After stacking 
the distributions of star particles projected to the observer's sky in the light-cone output 
and produced the stacked UV images for galaxies within the two brightest UV magnitude ranges, 
we find that the half light radii of the stacked images at $z \sim 7$ ($z \sim 8$) are 
$0.60~{\rm kpc}$ ($0.48~{\rm kpc}$) for the $M_{\rm UV} < -21$ cases and 
$0.55~{\rm kpc}$ ($0.55~{\rm kpc}$) for the $-21 < M_{\rm UV} < -20$ cases, respectively.
These results are consistent with the recent observational results \citep{Ono2013}. 

Finally, we have discussed a few additional issues derived from our simulation.
There is one candidate galaxy at $z = 11.8$ in the UDF \citep{Bouwens2011, Ellis2013, Oesch2013}.
In our simulation, we expect two galaxies brighter than the current 
detection limit and satisfying the color selection for $z > 10$ per one 
field-of-view of the UDF12 survey. 
Therefore, the number density is consistent with each other.
We also present the expected submm flux density of the mock-UDF12 galaxies.
The galaxies with $H_{160} \leq 27$, the expected submm flux density is $\geq 0.1$ mJy 
at 350 GHz in the observer's rest-frame.
This will be detectable with the ALMA.
We have examined the fraction of galaxies selected or not selected by the 
UDF12 survey. 
The detected and color-selected galaxies contribute to only $52\%, ~39\%, ~30\%$ and $12\%$ 
of the cosmic star formation rate density (SFRD) at $z \sim7, ~8, ~9$ and $z \sim 10$, respectively. 
A significant part (35--83\%) of the SFRD comes from galaxies satisfying the color selection 
criteria but not bright enough to be detected with the current detection limit. 
If we lower the detection limit of $< 32$ which corresponds to that of the JWST, 
the fraction of the detected and color-selected galaxies would drastically increases to 
$77\%, ~75\%, ~79\%$ and $72\%$ at $z \sim7, ~8, ~9$ and $z \sim 10$, respectively. 
Yet, $28 \%$ star forming activity at $z \sim 10$ remains undetectable even with the JWST. 
The fraction of the galaxies not satisfying the color criteria is as small 
as 5--14\% and they reside around not far from the color boundary. 
Therefore, we conclude that no galaxy with a very red UV color exists at $z > 7$ in our simulation.

\section*{Acknowledgments}
We thank an anonymous referee for her/his constructive comments. 
We also thank Tomoaki Ishiyama for his comments on the halo accretion rate of galaxies. 
Numerical simulations have been performed with the EUP, PRIMO 
and SGI cluster system installed 
at the Institute for the Physics and Mathematics of the Universe, 
University of Tokyo and with T2K Tsukuba at Centre for Computational
Sciences in University of Tsukuba. 
IS and AKI acknowledge the financial support of Grant-in-Aid for Young
Scientists (A: 23684010) by MEXT, Japan.
This work is partially supported by Grant-in-Aid for Young 
Scientists (S) (20674003) and 
by the FIRST program Subaru Measurements of Images and Redshifts (SuMIRe)
by the Council for Science and Technology Policy.
TO acknowledges the financial support  of Grant-in-Aid for Young Scientists (B: 24740112).

\bsp

\label{lastpage}

\end{document}